\documentclass{aa}

\usepackage{euclid}
\usepackage{graphicx}
\usepackage{natbib}
\usepackage{scalerel}
\usepackage{txfonts}
\usepackage[pdfencoding=auto,psdextra]{hyperref}
\hypersetup{
    colorlinks=true,
    linkcolor=blue,
    filecolor=magenta,      
    urlcolor=blue,
    citecolor=blue
}
\makeatletter
\renewcommand*\aa@pageof{, page \thepage{} of \pageref*{LastPage}}
\makeatother

\usepackage[utf8]{inputenc}

\usepackage[switch, modulo]{lineno}

\usepackage{amsmath}	
\usepackage{dsfont}
\usepackage{physics}
\usepackage{newtxtext,newtxmath}
\usepackage[export]{adjustbox}
\usepackage{tikz}
\usetikzlibrary{angles, quotes}
\usetikzlibrary{shapes.arrows,calc}
\usepackage{multirow}
\usepackage[margin=10pt,font=small,labelfont=bf]{caption} 

\newcommand{\figref}[1]{Fig. \ref{#1}} 
\newcommand{\secref}[1]{Sect. \ref{#1}}
\newcommand{\appref}[1]{Appendix \ref{#1}}
\newcommand{\Eq}[1]{Eq. (\ref{#1})}
\newcommand{\Eqbr}[1]{(Eq. \ref{#1})}

\renewcommand{\ee}{\mathrm{e}}
\newcommand{\btheta}{\boldsymbol{\theta}}
\newcommand{\bvartheta}{\boldsymbol{\vartheta}}
\newcommand{\ii}{\mathrm{i}}

\newcommand{\Omegam}{\Omega_{\mathrm{m}}}

\newcommand{\ttheta}[1]{\pmb{\theta}}

\newcommand{\gammac}{\gamma_{\mathrm{c}}}
\newcommand{\gammacs}[1]{\gamma_{\mathrm{c}{#1}}}

\newcommand{\Gammabarx}[2]{\overline{\Gamma}^{\times}_{#1}\left({#2}\right)}
\newcommand{\Norm}[2]{\mathcal{N}}

\newcommand{\gammattt}{\gamma_{\mathrm{t}\mathrm{t}\mathrm{t}}}
\newcommand{\gammaxtt}{\gamma_{\times\mathrm{t}\mathrm{t}}}
\newcommand{\gammatxt}{\gamma_{\mathrm{t}\times\mathrm{t}}}
\newcommand{\gammattx}{\gamma_{\mathrm{t}\mathrm{t}\times}}
\newcommand{\gammaxxt}{\gamma_{\times\times\mathrm{t}}}
\newcommand{\gammaxtx}{\gamma_{\times\mathrm{t}\times}}
\newcommand{\gammatxx}{\gamma_{\mathrm{t}\times\times}}
\newcommand{\gammaxxx}{\gamma_{\times\times\times}}

\newcommand{\gammat}{\gamma_{\mathrm{t}}}
\newcommand{\map}{\mathcal{M}_{\mathrm{ap}}}
\newcommand{\mperp}{\mathcal{M}_{\times}}
\newcommand{\mapthree}{\map^3}
\newcommand{\maptwomx}{\map^2 \mperp}
\newcommand{\mapmxtwo}{\map \mperp^2}
\newcommand{\mxthree}{\mperp^3}
\newcommand{\mapthreeens}{\left\langle\mapthree\right\rangle}
\newcommand{\maptwomxens}{\left\langle\maptwomx\right\rangle}
\newcommand{\mapmxtwoens}{\left\langle\mapmxtwo\right\rangle}

\newcommand{\mxthreeens}{\left\langle\mxthree\right\rangle}
\newcommand{\mapthreehat}{\hat{\mathcal{M}}_{\mathrm{ap}}^{3}}
\newcommand{\mathitq}{\mathrm{\textbf{\textit{q}}}}
\newcommand{\cqi}{\breve{\mathitq}_{i}}
\newcommand{\cqii}{\breve{\mathitq}_{i+1}}
\newcommand{\cqiii}{\breve{\mathitq}_{i+2}}
\newcommand{\cqa}{\breve{\mathitq}_1}
\newcommand{\cqb}{\breve{\mathitq}_2}
\newcommand{\cqc}{\breve{\mathitq}_3}
\newcommand{\cqac}{\breve{\mathitq}_1^*}
\newcommand{\cqbc}{\breve{\mathitq}_2^*}
\newcommand{\cqcc}{\breve{\mathitq}_3^*}

\newcommand{\StoN}{\mathrm{S/N}}
\newcommand{\binfunc}{\mathcal{B}}
\newcommand{\Ngal}{N_{\mathrm{gal}}}
\newcommand{\Npix}{N_{\mathrm{pix}}}
\newcommand{\NpixDelta}{\Npix^{(\Delta)}}
\newcommand{\npcf}{$N$PCF}
\newcommand{\treecorr}{{\small{\texttt{TREECORR}}} }
\newcommand*\circled[1]{\tikz[baseline=(char.base)]{
            \node[shape=circle,draw,inner sep=1pt] (char) {#1};}}

\tikzset{Pfeil/.style=%
{to path={let \p1 = ($(\tikztotarget)-(\tikztostart)$),
            \n1 = {int(mod(scalar(atan2(\y1,\x1))+360, 360))}, 
            \n2 = {veclen(\x1,\y1)}
        in \pgfextra{\typeout{\n1,\n2,\x1,\y1}}
         (\tikztotarget)
        node[draw,single arrow,
             minimum height=\n2-\pgflinewidth,
             inner sep=1ex,
             single arrow head extend=1ex,
             rotate=\n1, 
             anchor=tip, 
             ]{}
         }}}

\begin{document}

\defcitealias{Schneideretal2005}{SKL05}
\defcitealias{Asgarietal2021}{A21}

%
%
   \title{A roadmap to cosmological parameter analysis with third-order shear statistics III: Efficient estimation of third-order shear correlation functions and an application to the KiDS-1000 data}

\newcommand{\orcid}[1]{} 
\author{
Lucas Porth,$^{1}$\thanks{E-mail: lporth@astro.uni-bonn.de}
Sven Heydenreich,$^{1,2}$
Pierre Burger, $^{1}$
Laila Linke, $^{1,3}$
Peter Schneider$^{1}$}
\institute{$^{1}$University of Bonn, Argelander-Institut f\"ur Astronomie, Auf dem H\"ugel 71, 53121 Bonn, Germany\\
            $^{2}$ Department of Astronomy and Astrophysics, University of California, Santa Cruz, 1156 High Street, Santa Cruz, CA 95064 USA\\
            $^{3}$ Universit\"at Innsbruck, Institut f\"ur Astro- und Teilchenphysik, Technikerstr. 25/8, 6020 Innsbruck, Austria}
%
%
\abstract{
\textit{Context.} Third-order lensing statistics contain a wealth of cosmological information that is not captured by second-order statistics. However, the computational effort for estimating such statistics on forthcoming stage IV surveys is prohibitively expensive.\\
\textit{Aims.} We derive and validate an efficient estimation procedure for the three-point correlation function (3PCF) of polar fields such as weak lensing shear. We then use our approach to measure the shear 3PCF and the third-order aperture mass statistics on the KiDS-1000 survey.\\
\textit{Methods.} We construct an efficient estimator for third-order shear statistics which builds on the multipole decomposition of the 3PCF. We then validate our estimator on mock ellipticity catalogs obtained from $N$-body simulations. Finally, we apply our estimator to the KiDS-1000 data and present a measurement of the third-order aperture statistics in a tomographic setup.\\
\textit{Results.} Our estimator provides a speedup of a factor of $\sim \! 100$\,-\,$1000$ compared to the state-of-the-art estimation procedures. It is also able to provide accurate measurements for squeezed and folded triangle configurations without additional computational effort. We report a significant detection of the tomographic third-order aperture mass statistics in the KiDS-1000 data $(\StoN=6.69)$. \\
\textit{Conclusions.} Our estimator will make it computationally feasible to measure third-order shear statistics in forthcoming stage IV surveys. Furthermore, it can be used to construct empirical covariance matrices for such statistics. 
    }
%
%
\keywords{gravitational lensing: weak – methods: numerical – large-scale structure of Universe}
%
%
   \titlerunning{Efficient Estimation of the Shear 3PCF}
   \authorrunning{L. Porth et al.}
   \maketitle
%
%
%
%
   
\section{Introduction}
The most widely used set of statistics for extracting cosmological information from galaxy or weak gravitational lensing surveys is related to $N$-point correlation functions (\npcf s). Estimating those quantities from the position or shapes of a set of observed tracers, one reveals parts of the statistical properties of the underlying cosmic large-scale structure. In particular, for weak gravitational lensing surveys the analysis of shear $2$PCFs and its harmonic analog, the power spectrum, have made it possible to put tight constraints on the matter clustering parameter $S_8 \equiv \sigma_8 \sqrt{\Omegam/0.3}$, where $\sigma_8$ is the normalization of the matter power spectrum and $\Omegam$ is the matter density parameter \citep{Heymansetal2013,Hildebrandtetal2017,Troxeletal2018,Hikageetal2019,Hamanaetal2020,Asgarietal2021,Amonetal2022,Seccoetal2022,Dalaletal2023,Lietal2023}. 

While most of the theoretical and observational work has been done at the second-order level, it is well known that through this choice one is sensitive only to the Gaussian information of the density field. To extract additional information including contributions from non-linear structure formation, one needs to resort to other higher-order statistics, some of which are related to higher-order moments \citep{JainSeljak1997,Schneideretal1998,Vincanzaetal2018,Gattietal2020}, peaks \citep{KruseSchneider1999,Hamanaetal2004,HarnoisDerapsetal2021}, the one-point probability distribution function \citep{Pattonetal2017,Barthelemyetal2020,Boyleetal2021,Giblinetal2023}, or the topological properties \citep{Kratochviletal2012,Parronietal2020,Heydenreichetal2022b} of the density field. Besides the gain in cosmological information, it has been shown that when performing a joint analysis together with second-order statistics, higher-order statistics can also break degeneracies between parameters of cosmological,  astrophysical or systematic origin and therefore have the potential to yield much tighter cosmological constraints \citep{KilbingerSchneider2005,TakadaJain2003, Sembolonietal2008,Pyneetal2021}. 

Higher-order statistics from the correlation function hierarchy consist of the $(N>2)$-point correlation functions, from which the three-point correlation function (3PCF) is the most prominent example \citep{SchneiderLombardi2003,ZaldarriagaScoccimarro2003}. Besides the challenges to accurately model the cosmological and observational effects on the shape of the 3PCF \citep{TakadaJain2003,Heydenreichetal2022}, it is also computationally much more expensive to estimate them, compared to the 2PCF. A naive estimator would require putting all triplets of tracers into a predefined set of bins describing the triplets'  configuration. The resulting computational complexity for a survey of $\Ngal$ tracers is $\mathcal{O}(\Ngal^3)$, rendering this brute-force approach computationally prohibitive. To facilitate an estimation of the 3PCF of scalar (spin-0) and polar (spin-2) observables for current surveys, $k$d-tree based methods have been developed \citep{Jarvisetal2004,ZhangPen2005,Kilbingeretal2014} and applied to a wide range of data sets \citep{Bernardeauetal2002,Sembolonietal2011,Fuetal2014,Seccoetal2022}. Even though the computational speedup compared to the brute-force estimator is remarkable, it is most likely still too time-consuming to apply it to stage IV cosmic shear surveys like \textit{Euclid} \citep{Laureijsetal2011} or the Vera Rubin Observatory Legacy Survey of Space and Time \citep{Ivezicetal2019}. A different class of estimators has been tailored towards the multipole components of the \npcf s of scalar fields \citep{ChenSzapudi2005,SlepianEisenstein2016,Philcoxetal2022a}. As the multipoles can be estimated in quadratic time complexity for a discrete galaxy catalog, or with an $\Npix \log(\Npix)$ scaling for data distributed on a grid of $\Npix$ pixels, this estimation strategy makes it feasible to efficiently and accurately measure higher-order correlation functions on large data sets. For example, the first measurement of the 4PCF of the scalar galaxy position field was presented in \cite{Houetal2022a} and \cite{Philcoxetal2022b}.

In this work, we extend the multipole decomposition of the scalar 3PCF to the natural components of the 3PCF associated with spin-2 tracer fields, such as weak lensing shear. Additionally, we construct an efficient estimator that adaptively switches between a discrete and grid-based prescription for triangle sides of different lengths. With those modifications, we can measure the full 3PCF for any triangle configuration on a stage-III survey in a few CPU hours without significantly sacrificing accuracy compared to the brute-force estimator. 

This work belongs to a series of papers that aim for a cosmological parameter analysis using third-order shear statistics. \cite{Heydenreichetal2022} introduces an analytical model for the third-order aperture statistics based on the Bihalofit formula \citep{Takahashietal2020} for the matter bispectrum. In \cite{Linkeetal2023}, an analytical covariance matrix for the third-order aperture statistics is derived in the presence of a finite survey extent. \cite{Burgeretal2023} extends the analytical model to a tomographic setting, includes methods to account for the impact of intrinsic alignments and baryonic feedback, and presents a cosmological analysis using a joint data vector of the COSEBIs and the third-order aperture mass statistics measured in the KiDS-1000 data.

This paper is structured as follows. In \secref{sec:ThirdOrderMeasuresOfCosmicShear}, we introduce the shear 3PCF and their natural components $\Gamma_\mu$. In \secref{sec:Estimator}, we derive our estimator for the $\Gamma_\mu$ and discuss some approximations. In \secref{sec:Validation}, we validate our estimator on a large suite of N-body simulations, the SLICS ensemble, and compare it with measurements obtained with \treecorr \citep{Jarvisetal2004}. In \secref{sec:ApertureMass} we introduce the third-order aperture measures, show how they can be obtained from our shear 3PCF estimator and validate our implementation. In \secref{sec:KiDSMeasurement}, we apply our estimator to the fourth data release of the Kilo Degree Survey (\citealt{Kuijkenetal2019}, hereafter KiDS-1000). After obtaining an empirical estimate of the expected correlation matrix for the KiDS-1000 data from a large suite of mock catalogs we present our measurement of the third-order aperture statistics in the KiDS-1000 data. We conclude in \secref{sec:Conclusions}.


\section{Third-order measures of cosmic shear}\label{sec:ThirdOrderMeasuresOfCosmicShear}
This section aims to introduce the main concepts related to weak gravitational lensing with an emphasis on the third-order shear correlation functions; for extensive reviews of the weak gravitational lensing see \cite{BartelmannSchneider2001,Kilbinger2015,Dodelson2017,Mandelbaum2018}. In order to keep the notation concise, we do not consider tomographic setups in this and the following section but collect the corresponding expressions in \secref{ssec:TomoEstimator}. Furthermore, we do not consider higher-order effects like reduced shear, source redshift clustering, or astrophysical effects, such as intrinsic alignments.
\subsection{Basic notions}
The two fundamental quantities in weak gravitational lensing are the convergence $\kappa$ and the shear $\gamma$, which describe the isotropic stretching and distortion experienced by a light bundle propagating through spacetime. The convergence $\kappa$ can be written as a weighted projection of the density contrast $\delta$ along the line of sight 
\begin{align}\label{eq:kappadef}
    \kappa(\btheta) = \int_0^{\chi_{\mathrm{max}}} \dd \chi' W(\chi') \  \delta\left[f_K(\chi')\btheta;\chi'\right] \ ,
\end{align}
where the associated projection kernel $W$ is defined as 
\begin{align}
    W(\chi) \equiv \frac{3\Omegam H_0^2}{2c^2} \frac{f_K(\chi)}{a(\chi)} \int_{z(\chi)}^{\infty} \dd z' n(z')  \frac{f_K[\chi(z')-\chi]}{f_K[\chi(z')]} 
\end{align}
and $f_K[\chi(z)]$ denotes the comoving angular diameter distance of the comoving radial distance $\chi$ at redshift $z$ and we introduced the dimensionless matter density parameter $\Omegam$ and the Hubble constant $H_0$, as well as the scale factor $a$ and the source redshift distribution $n(z)$. For the remainder of this work, we assume a flat universe for which $f_K(\chi) = \chi$.

The components of the complex shear field $\gamma$ can be characterised with respect to a Cartesian coordinate frame, $\gammac \equiv \gamma_1 + \ii \gamma_2$. To evaluate the shear at position $\btheta$ in a reference frame, that is rotated by an angle $\zeta$ with respect to the Cartesian basis one has
\begin{align}\label{eq:gammatx_def}
    \gamma(\btheta;\zeta) \equiv -\left[\gammat(\btheta;\zeta) + \ii \gamma_\times(\btheta;\zeta)\right] \equiv -\gammac(\btheta) \ \ee^{-2\ii\zeta} \ ,
\end{align}
where we introduced the tangential $(\gammat)$ and cross-components $(\gamma_\times)$ of the shear with respect to the projection direction $\zeta$.
\subsection{The shear 3PCF and its natural components}
Due to the polar nature of the shear, there are $8$ different real-valued components for its three-point correlator. \cite{SchneiderLombardi2003} regroup those components into four complex-valued \textit{natural} components $\Gamma_\mu$, that do not mix and only transform by some phase factor under rotations of the corresponding triangle. Parameterizing a triangle as depicted in \figref{fig:TriangleNotation} and choosing some arbitrary projection $\boldsymbol{\mathcal{P}}$ in which the individual shear components are rotated by angles $\zeta_i$, we define the natural components of the shear 3PCF as\footnote{We note that our definition of the $\Gamma_\mu^{ \boldsymbol{\mathcal{P}}}$ is different from the one found in \cite{SchneiderLombardi2003}, but instead matches the configuration of the third-order correlators introduced in \cite{Jarvisetal2004} and the \treecorr implementation.}
\begin{align}
    \label{eq:Gamma0_def}
    \Gamma_0^{ \boldsymbol{\mathcal{P}}} (\vartheta_1, \vartheta_2, \phi) &= 
    \left\langle \ 
    \gamma\left(\mathbf{X_1};\zeta_1\right) \ \gamma\left(\mathbf{X_2};\zeta_2\right) \ 
    \gamma\left(\mathbf{X_3};\zeta_3\right)
    \ \right\rangle \ , \\
    \Gamma_1^{ \boldsymbol{\mathcal{P}}} (\vartheta_1, \vartheta_2, \phi) &= 
    \left\langle \ \gamma^*\hspace{-.1cm}\left(\mathbf{X_1};\zeta_1\right) \ \gamma\left(\mathbf{X_2};\zeta_2\right) \ 
    \gamma\left(\mathbf{X_3};\zeta_3\right)
    \ \right\rangle \ , \\
    \Gamma_2^{ \boldsymbol{\mathcal{P}}} (\vartheta_1, \vartheta_2, \phi) &= 
    \left\langle \ 
    \gamma\left(\mathbf{X_1};\zeta_1\right) \ \gamma^*\hspace{-.1cm}\left(\mathbf{X_2};\zeta_2\right) \ \gamma\left(\mathbf{X_3};\zeta_3\right) \ 
    \right\rangle \ , \\
    \label{eq:Gamma3_def}
    \Gamma_3^{ \boldsymbol{\mathcal{P}}} (\vartheta_1, \vartheta_2, \phi) &= 
    \left\langle \ 
    \gamma\left(\mathbf{X_1};\zeta_1\right) \  \gamma \left(\mathbf{X_2};\zeta_2\right) \ \gamma^*\hspace{-.1cm} \left(\mathbf{X_3};\zeta_3\right)
    \ \right\rangle \ ,
\end{align}
where $\vartheta_i \equiv |\bvartheta_i|$ and we have assumed that the projection directions $\zeta_i$ are defined in terms of the vertices of the triangle; this implies that the natural components are invariant under rotations of the triangle such that a parametrization in terms of three variables is sufficient. We note that any of the natural components can be written in terms of the elements
\begin{align}
    \label{eq:3PCF_rawcomponents}
     \{
    \gammattt, 
    \gammaxtt, \gammatxt, \gammattx, 
    \gammatxx, \gammaxtx, \gammaxxt, 
    \gammaxxx 
    \} \ ,
\end{align}
 where we introduced shorthand notation like $\gammattt \equiv \left\langle\gammat \gammat \gammat\right\rangle$ for the various triple products of shears \citep{SchneiderLombardi2003}.
Inverting the relations \eqref{eq:Gamma0_def}--\eqref{eq:Gamma3_def} one can relate the components in \Eq{eq:3PCF_rawcomponents} to the natural components.

\subsection{Choice of projections for the shear 3PCF}
\begin{figure}
  \centering
  \includegraphics[width=.9\linewidth]{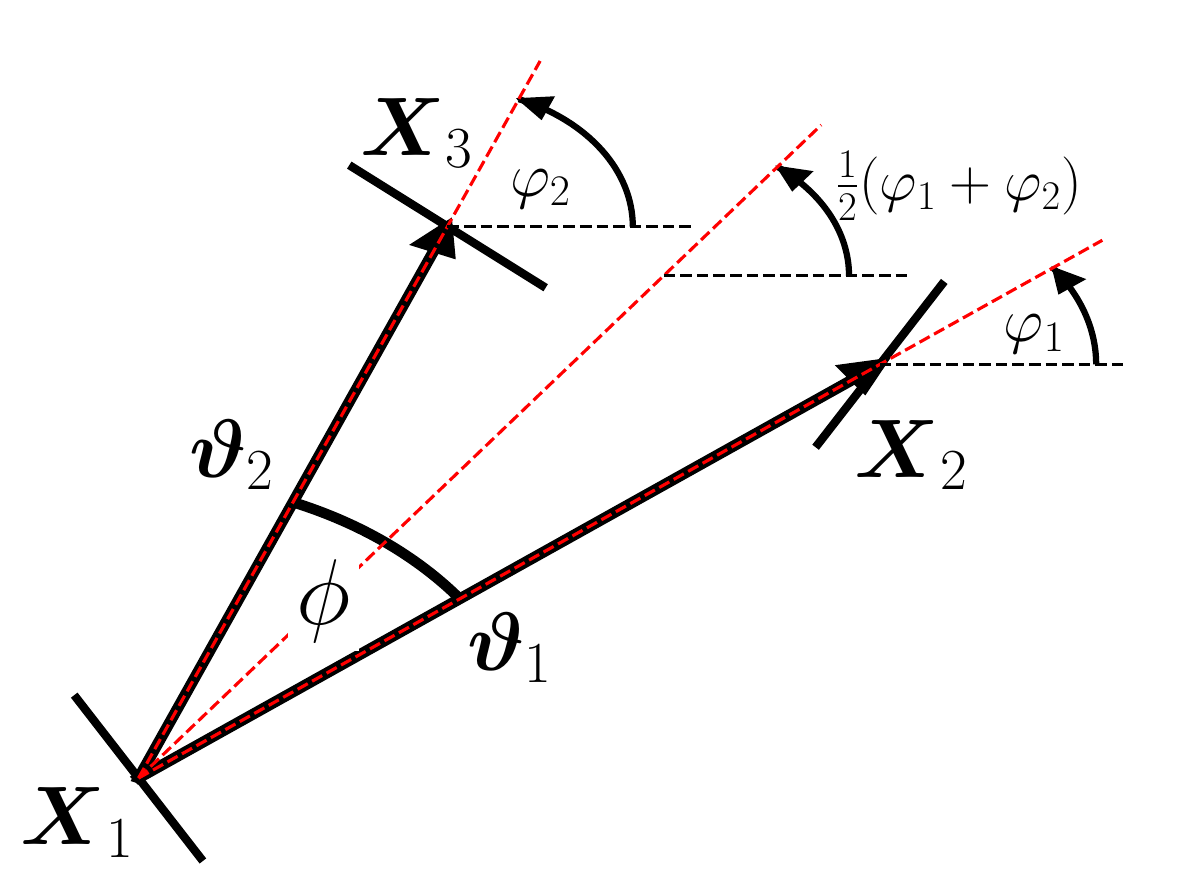}\\[-0cm]
  \includegraphics[width=.9\linewidth]{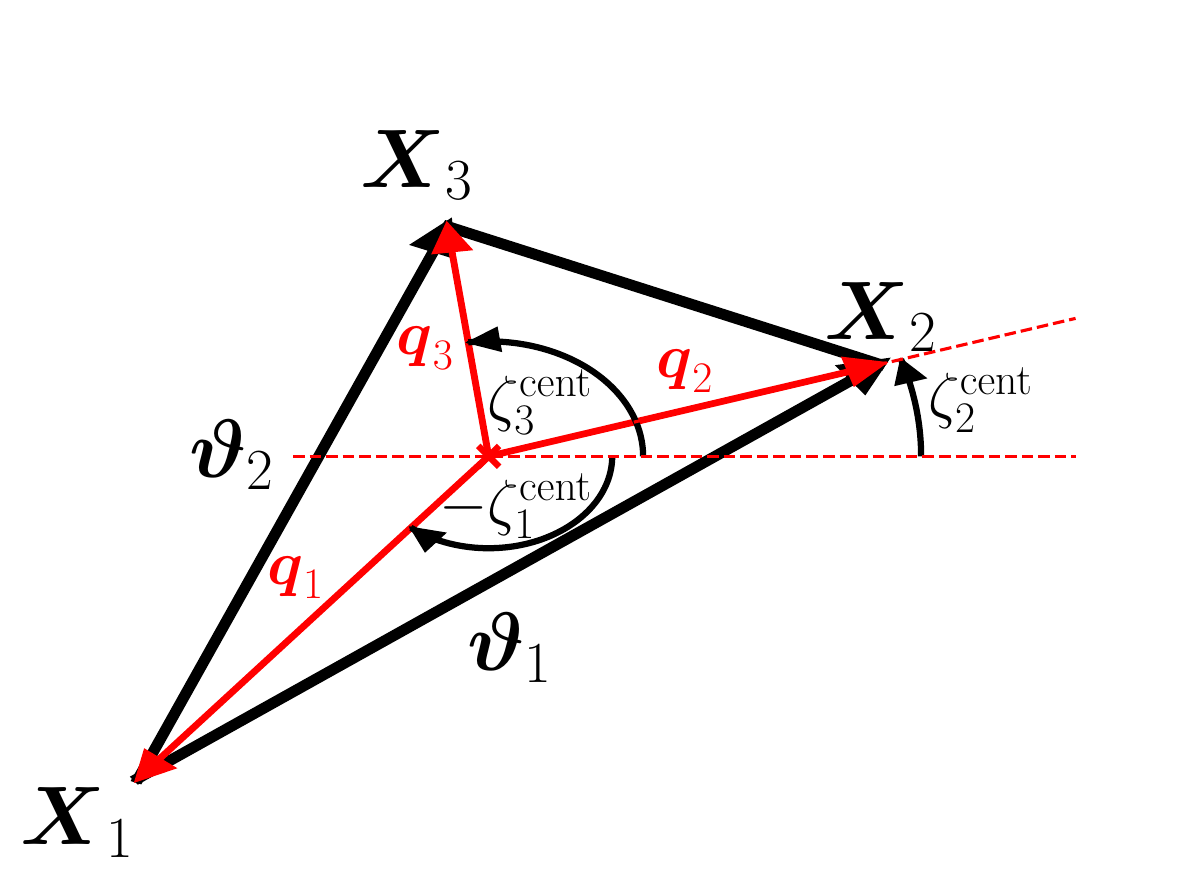}
\caption{\textit{Upper panel}: Parametrization of a triplet of shears used in this work. For some shear at position $\boldsymbol{X}_1$, we denote the connecting lines to the two other shear as $\boldsymbol{\vartheta}_i$ and the enclosing angle as $\phi$. The red dashed lines show the directions of the $\times$ projection \Eq{eq:xprojection} for which the three projection axes intersect in $\boldsymbol{X}_1$. \textit{Lower panel}: Definition of the geometric quantities involved in the centroid projection \Eq{eq:centroidprojection}. We see that the $\boldsymbol{q}$-vectors connect the centroid of the triangle with the triangles' vertices.}
\label{fig:TriangleNotation}
\end{figure}

For this work, we employ two particular projections. The first of them, which we dub the $\times$-projection, is defined by the rotation angles
\begin{align}\label{eq:xprojection}
    \zeta_1^\times = \frac{1}{2} 
    \left(\varphi_1+\varphi_2\right)
    \ \ \ , \ \ \ 
    \zeta_2^\times = \varphi_1
    \ \ \ , \ \ \ 
    \zeta_3^\times = \varphi_2\
     \ 
\end{align}
and the corresponding projection axes are depicted in the upper panel of \figref{fig:TriangleNotation}. As we will see later, this projection turns out to be useful for the estimator derived in \secref{sec:Estimator}. For the second projection shown in the lower panel of \figref{fig:TriangleNotation}, we first define the centroid of the triangle as the intersection of its three medians and then project the shear at position $\mathbf{X}_i$ along the direction $\mathitq_i$ of the median passing through $\mathbf{X}_i$:
\begin{align}\label{eq:centroidprojection}
    \ee^{2\ii\zeta_1^{\mathrm{cent}}} = \frac{\cqa}{\cqac}
    \ \ \ , \ \ \ 
    \ee^{2\ii\zeta_2^{\mathrm{cent}}} =  \frac{\cqb}{\cqbc}
    \ \ \ , \ \ \ 
    \ee^{2\ii{\zeta_3^{\mathrm{cent}}}} = \frac{\cqc}{\cqcc} \  ,
\end{align}
where the $\mathitq_i$ are given as
\begin{align}\label{eq:QvecDef}
    \mathitq_1 = -\frac{\bvartheta_1+\bvartheta_2}{3}
    \ \ \ , \ \ \
    \mathitq_2 = \frac{2\bvartheta_1-\bvartheta_2}{3}
    \ \ \ , \ \ \
    \mathitq_3 = \frac{2\bvartheta_2-\bvartheta_1}{3}    
\end{align}
and we defined $\cqi \equiv q_{i,1} + \ii q_{i,2}$. The corresponding $\Gamma_{\mu}^{\mathrm{cent}}$ are useful for connecting the shear 3PCF to the third-order aperture statistics and is being used by other estimators for the 3PCF such as \treecorr.

We can convert between the two projections using \Eq{eq:gammatx_def}. For example, the zeroth natural component in the $\times$-projection can be written as
\begin{align*}
    \Gamma^\mathrm{\times}_0 (\vartheta_1,\vartheta_2,\phi)
    &=
    -\langle \gammac(\mathbf{\theta})\gammac(\mathbf{\theta}+\mathbf{\vartheta}_1)\gammac(\mathbf{\theta}+\mathbf{\vartheta}_2) \rangle \ \ee^{-3\ii(\varphi_1+\varphi_2)}
    \nonumber \\
    &= \Gamma_0^{\mathrm{cent}}(\vartheta_1,\vartheta_2,\phi) \frac{\left| \cqa \cqb \cqc \right|^2}{\left( \cqac \cqbc \cqcc \right)^2} \  \ee^{-3\ii(\varphi_1+\varphi_2)} \ ,
\end{align*}
from which one can immediately read off $\Gamma_0^{\mathrm{cent}}$. Repeating similar computations for the other components and using that $\varphi_2 = \varphi_1 + \phi$, we arrive at
\begin{align}\label{eq:X2Centroid0}
    \Gamma_0^{\mathrm{cent}}(\vartheta_1,\vartheta_2,\phi)
    &=
\Gamma^\mathrm{\times}_0(\vartheta_1,\vartheta_2,\phi)
    \frac{\cqac}{\cqa}
    \frac{\cqbc}{\cqb}
    \frac{\cqcc}{\cqc}
    \ee^{6\ii\varphi_1}\ee^{3\ii\phi} 
    \ , \\
    \Gamma_1^{\mathrm{cent}}(\vartheta_1,\vartheta_2,\phi)
    &=
    \Gamma^\mathrm{\times}_1(\vartheta_1,\vartheta_2,\phi)
    \frac{\cqa}{\cqac}
    \frac{\cqbc}{\cqb}
    \frac{\cqcc}{\cqc}
    \ee^{2\ii\varphi_1}\ee^{\ii\phi}
    \ , \\
    \Gamma_2^{\mathrm{cent}}(\vartheta_1,\vartheta_2,\phi)
    &=
    \Gamma^\mathrm{\times}_2(\vartheta_1,\vartheta_2,\phi)
    \frac{\cqac}{\cqa}
    \frac{\cqb}{\cqbc}
    \frac{\cqcc}{\cqc}
    \ee^{2\ii\varphi_1}\ee^{3\ii\phi}
    \ , \\
    \Gamma_3^{\mathrm{cent}}(\vartheta_1,\vartheta_2,\phi)
    &=
    \Gamma^\mathrm{\times}_3(\vartheta_1,\vartheta_2,\phi)
    \frac{\cqac}{\cqa}
    \frac{\cqbc}{\cqb}
    \frac{\cqc}{\cqcc}
    \ee^{2\ii\varphi_1}\ee^{-\ii\phi} \ ,
    \label{eq:X2Centroid3}
\end{align}
where the rotation angles can be expressed as
\begin{align*}
    \frac{\cqac}{\cqa}\ee^{2\ii\varphi_1}
    &=
    \frac{\vartheta_1+\vartheta_2\ee^{-\ii\phi}}{\vartheta_1+\vartheta_2\ee^{\ii\phi}}
    \ , \hspace{.5cm}
    \frac{\cqbc}{\cqb}\ee^{2\ii\varphi_1} 
    = 
    \frac{\vartheta_2\ee^{-\ii\phi}-2\vartheta_1}{\vartheta_2\ee^{\ii\phi}-2\vartheta_1}
    \ , \\
    \frac{\cqcc}{\cqc}\ee^{2\ii\varphi_1}
    &=
    \frac{\vartheta_1-2\vartheta_2\ee^{-\ii\phi}}{\vartheta_1-2\vartheta_2\ee^{\ii\phi}}
    \ .
\end{align*}
\subsection{Bin-averaged shear 3PCF}
When estimating the $\Gamma_\mu^{\mathcal{P}}$ from a finite set of galaxy ellipticities, one does not have direct access to every possible triangle shape but instead collects the point triplets into radial and angular bins. Such a measurement then provides an estimator for the bin-averaged shear 3PCF $\overline{\Gamma}_\mu^{\mathcal{P}}$:
\begin{align}\label{eq:Gamma0_binned}
    \overline{\Gamma}_\mu^{\mathcal{P}}(&\Theta_i,\Theta_j,\Phi_k)  \nonumber \\
   &\equiv \int_{\Theta_i}\frac{\dd \vartheta_i \ \vartheta_i}{A_i} \int_{\Theta_j}\frac{\dd \vartheta_j \ \vartheta_j}{A_j} 
   \int_{\Phi_k}\frac{\dd \phi}{|\Delta\Phi_k|} \ 
\Gamma_\mu^{\mathcal{P}}(\vartheta_i,\vartheta_j,\phi) \ ,
\end{align}
where for each radial bin $\Theta_i$ we have $\vartheta_i \in [\Theta_{\mathrm{low},i},\Theta_{\mathrm{up},i}]$ and we define $A_i\equiv (\Theta_{\mathrm{up},i}^2-\Theta_{\mathrm{low},i}^2)/2$. Similarly, for the angular bins we have $\phi \in [\Phi_{\mathrm{low},k},\Phi_{\mathrm{up},k}]$ and $|\Delta\Phi_k|\equiv \Phi_{\mathrm{up},k}-\Phi_{\mathrm{low},k}$.

\section{An efficient estimator for the shear 3PCF}\label{sec:Estimator}
Given a weak lensing catalog consisting of $\Ngal$ galaxies at positions $\btheta_i$ with ellipticities\footnote{In this work, we assume that the measured ellipticity of a galaxy provides an unbiased, albeit noisy, estimate of the underlying shear field at the galaxies' angular position.} $\gammacs{,i}$ and weights $w_i$, the standard estimator for the bin-averaged natural components of the shear 3PCF consists of assigning all galaxy triplets to their corresponding triangle configuration bin and then averaging over those. In particular, the estimator for the zeroth natural component using the $\times$-projection \Eq{eq:xprojection} becomes
\begin{align}\label{eq:3PCF_tripletestimator}
\hat{\overline{\Gamma}}^{\times}_0(\Theta_1,\Theta_2,\Phi)
\equiv \frac{\Upsilon_0^\times(\Theta_1,\Theta_2,\Phi) }{\mathcal{N}(\Theta_1,\Theta_2,\Phi) } \ ,
\end{align}
where
\begin{align}
\label{eq:TripletUpsilon}
\Upsilon_0^\times(\Theta_1,\Theta_2,\Phi)  
    &\equiv 
    -\sum_{i,j,k=1}^{\Ngal}w_i\gammacs{,i} \ w_j\gammacs{,j}\  w_k\gammacs{,k} \ 
    \ee^{-3\ii(\varphi_{ij}+\varphi_{ik})}
    \nonumber \\ &\times
    \binfunc\left(\theta_{ij}\in\Theta_1\right) \ 
    \binfunc\left(\theta_{ik}\in\Theta_2\right)
    \binfunc\left(\phi_{ijk}\in\Phi\right) \ ,
    \\
\label{eq:TripletNorm}
\mathcal{N}(\Theta_1,\Theta_2,\Phi)  
    &\equiv 
    \sum_{i,j,k=1}^{\Ngal}w_i \ w_j\  w_k\ 
    \nonumber \\ &\times
    \binfunc\left(\theta_{ij}\in\Theta_1\right) \ 
    \binfunc\left(\theta_{ik}\in\Theta_2\right)
    \binfunc\left(\phi_{ijk}\in\Phi\right) \ .
\end{align}
In the preceding equations we defined $\theta_{ij} \equiv |\btheta_{ij}| \equiv |\btheta_j-\btheta_i|$ and $\phi_{ijk}\equiv\varphi_{ik}-\varphi_{ij}$, in with $\varphi_{ij}$ denoting the polar angle of $\btheta_{ij}$. We furthermore introduced the bin selection function $\binfunc(x \in X)$ which is unity if $x\in X$ and zero otherwise; the different combinations of the bin selection functions then specify the various triangle configurations.
\subsection{Multipole decomposition}\label{ssec:MultipoleDecomposition}
Let us first consider the numerator in \Eq{eq:3PCF_tripletestimator}. Instead of a discrete angular binning, one can instead perform a multipole decomposition of $\Upsilon_0^\times$. For an infinitely dense angular binning, one can decompose
\begin{align}\label{eq:Upsilon0Decomposition}
    \Upsilon_0^\times(\Theta_1,\Theta_2,\phi)
    \equiv 
    \frac{1}{2\pi} \sum_{n=-\infty}^\infty \Upsilon_{0,n}^\times(\Theta_1,\Theta_2) \ \ee^{\ii n\phi} \ ,
\end{align}
in which the $n$th multipole can be successively manipulated to yield
\begin{align}\label{eq:Upsilon0_Derivation}
    \Upsilon_{0,n}^\times&(\Theta_1,\Theta_2)
    \nonumber \\ &\equiv
    \int_0^{2\pi} \dd \phi \ \Upsilon_0^\times(\Theta_1,\Theta_2,\phi) \ee^{-\ii n\phi}
    \nonumber \\ &=
    - \int_0^{2\pi} \dd \phi \sum_{i,j,k=1}^{\Ngal}w_i\gammacs{,i} \ w_j\gammacs{,j}\  w_k\gammacs{,k} \ 
    \ee^{-3\ii(\varphi_{ij}+\varphi_{ik})}
    \nonumber \\ &\hspace{1cm}\times
    \binfunc\left(\theta_{ij}\in\Theta_1\right) \ 
    \binfunc\left(\theta_{ik}\in\Theta_2\right) \
    \binfunc\left(\phi_{ijk}\in\{\phi\}\right)
    \ \ee^{-\ii n\phi}
    \nonumber \\ &=
    -\sum_{i,j,k=1}^{\Ngal}w_i\gammacs{,i} \ w_j\gammacs{,j}\  w_k\gammacs{,k} \ 
    \ee^{-3\ii(\varphi_{ij}+\varphi_{ik})} \ee^{-\ii n(\varphi_{ik}-\varphi_{ij})}
    \nonumber \\ &\hspace{1cm}\times
    \binfunc\left(\theta_{ij}\in\Theta_1\right) \ 
    \binfunc\left(\theta_{ik}\in\Theta_2\right)
    \nonumber \\ &=
    -\sum_{i=1}^{\Ngal}w_i\gammacs{,i} \ \ \ 
    \sum_{j=1}^{\Ngal}w_j\gammacs{,j} \ \ee^{\ii(n-3)\varphi_{ij}} \ \binfunc\left(\theta_{ij}\in\Theta_1\right)
    \nonumber \\ &\hspace*{2.1cm}\times
    \sum_{k=1}^{\Ngal}w_k\gammacs{,k} \ \ee^{-\ii(n+3)\varphi_{ik}} \ \binfunc\left(\theta_{ik}\in\Theta_2\right)
    \nonumber \\ &\equiv
    -\sum_{i=1}^{\Ngal}w_i\gammacs{,i} \ 
    G_{n-3}^{\mathrm{disc}}(\btheta_i;\Theta_1) \ G_{-n-3}^{\mathrm{disc}}(\btheta_i;\Theta_2) \ ,
\end{align}
where in the second step we inserted \Eq{eq:TripletUpsilon} and in the final step defined the quantity
\begin{align}\label{eq:Gn_disc}
    G_{n}^{\mathrm{disc}}(\btheta_i;\Theta)
    &=
    \sum_{k=1}^{\Ngal}w_k\gammacs{,k} \ \ee^{\ii n\varphi_{ik}} \ \binfunc\left(\theta_{ik}\in\Theta \right) \nonumber \\
    &\equiv 
    \sum_{k=1}^{\Ngal}w_k\gammacs{,k} \ g_{n;\Theta}(\btheta_{ik})
    \ .
\end{align}
Repeating similar calculations for the numerators of the other three natural components we find 
\begin{align}
    \Upsilon_{1,n}^\times(\Theta_1,\Theta_2) &= 
    -\sum_{i=1}^{\Ngal}w_i\gammacs{,i}^* \ 
    G_{n-1}^{\mathrm{disc}}(\btheta_i;\Theta_1) \ G_{-n-1}^{\mathrm{disc}}(\btheta_i;\Theta_2) 
    \label{eq:Upsilon1}
    \ , \\
    \Upsilon_{2,n}^\times(\Theta_1,\Theta_2) &= 
    -\sum_{i=1}^{\Ngal}w_i\gammacs{,i} \ 
    \left(G_{-n-1}^{\mathrm{disc}}(\btheta_i;\Theta_1)\right)^* \ G_{-n-3}^{\mathrm{disc}}(\btheta_i;\Theta_2) 
    \label{eq:Upsilon2}
    \ , \\
    \Upsilon_{3,n}^\times(\Theta_1,\Theta_2) &= 
    -\sum_{i=1}^{\Ngal}w_i\gammacs{,i} \ 
    G_{n-3}^{\mathrm{disc}}(\btheta_i;\Theta_1) \ \left(G_{n-1}^{\mathrm{disc}}(\btheta_i;\Theta_2)\right)^*
    \ ,
    \label{eq:Upsilon3}
\end{align}
where for the final two equations we used
\begin{align*}
    \left(G_{-n}^{\mathrm{disc}}(\btheta_i;\Theta)\right)^*
    =
    \sum_{k=1}^{\Ngal}w_k\gammacs{,k}^* \ g_{n;\Theta}(\btheta_{ik}) \ .
\end{align*}
For the remainder of this work, we only write down the $\mu=0$ component for any equation derived from the $\Upsilon_{\mu}$; the expressions for the other components can be inferred from Eqs. \eqref{eq:Upsilon1}-\eqref{eq:Upsilon3}.

Now switching to the denominator \Eq{eq:TripletNorm} we can perform a derivation similar to the one leading to \Eq{eq:Upsilon0_Derivation} to find the multipole expansion
\begin{align}\label{eq:NormDecomposition}
    \mathcal{N}(\Theta_1,\Theta_2,\phi)
    \equiv 
    \frac{1}{2\pi} \sum_{n=-\infty}^\infty \mathcal{N}_{n}(\Theta_1,\Theta_2) \ \ee^{\ii n\phi} \ ,
\end{align}
with multipole components
\begin{align}\label{eq:3PCF_Normalization}
    \mathcal{N}_n(\Theta_1,\Theta_2) &= 
    \sum_{i=1}^{\Ngal}w_i \ 
    W_{n}^{\mathrm{disc}}(\btheta_i;\Theta_1) \ \left(W_{n}^{\mathrm{disc}}(\btheta_i;\Theta_2)\right)^* \ ,
    \\
    W_{n}^{\mathrm{disc}}(\btheta_i;\Theta)
     &\equiv 
    \sum_{k=1}^{\Ngal} \ w_k \ g_{n;\Theta}(\btheta_{ik}) \ .
\end{align}
The decomposition of the denominator has already been explored in the context of galaxy clustering in which the corresponding 3PCF is a scalar \citep{SlepianEisenstein2016,Philcoxetal2022a}. Here we have shown that using the $\times$-projection \Eq{eq:xprojection} gives structurally equivalent expressions for the multipole components of the 3PCF of polar fields. 

Regarding the computational complexity, we see that each multipole in the decompositions of the $\Upsilon_\mu^\times$ and $\mathcal{N}$ involves a double sum over the whole ellipticity catalog making the estimation process of the shear 3PCF formally scale as $\mathcal{O}\left(\Ngal^2\right)$. In practice, the largest separation $\theta_{\mathrm{max}}$ for which we want to compute the 3PCF might be much smaller than the survey extent; in this case, one can employ an efficient search routine that readily finds the various galaxies located in the rings around the $\btheta_i$. Assuming a constant runtime\footnote{In our implementation we use a spatial hashing data structure, as shown i.e. in \cite{Porthetal2021} this method does indeed scale as $\mathcal{O}(1)$.} for this search algorithm the scaling is brought down to  $\mathcal{O}\left(\Ngal 
 \ \overline{n} \theta_{\mathrm{max}}^2\right)$.
\subsection{Grid-based approximation}\label{ssec:GridApproximation}
\begin{figure}
  \centering
  \includegraphics[width=.9\linewidth]{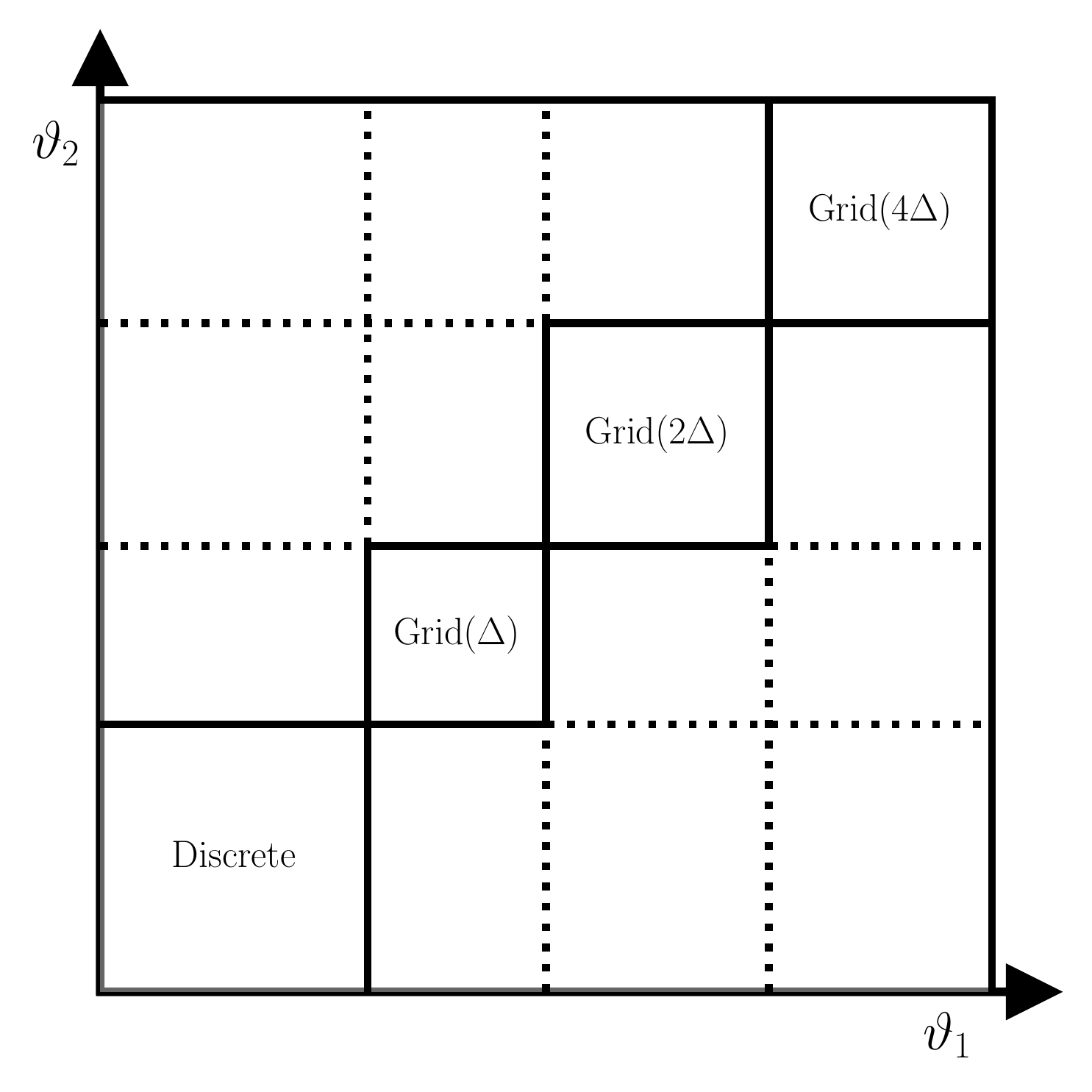}
\caption{General strategy for computing the three-point correlators used in this work. For small angular scales, we use the prescription for discrete data to compute the $G_n$ and the $\Upsilon^\times_{i,n}$ while for larger scales, we use the grid-based approximation with varying resolution scales $\Delta$. For the off-diagonal elements, we first compute the $G_n$ of each radial bin using the corresponding method and then map them to the grid of the larger resolution scale from which the $\Upsilon^\times_{i,n}$ are obtained.}
\label{fig:EstimatorGeneralStrategy}
\end{figure}
While the time complexity of the multipole estimator derived in \secref{ssec:MultipoleDecomposition} provides a significant speedup compared to the naive estimator and is competitive with tree-based implementations, the quadratic scaling is still prohibitive for applications in stage-IV surveys. However, we note that most of the complexity stems from the largest separations for which neighbouring galaxies make similar contributions to the $G_n^{\mathrm{disc}}$ and $W_n^{\mathrm{disc}}$.\footnote{As an example, the exponential factors appearing in the $G_n$ and $W_n$ consist of $2n$ extrema, such that for a radial bin with lower bound $\Theta_\mathrm{low}$ galaxies separated by $d \ll \Theta_\mathrm{low}/n$ will give similar contributions for the first $n$ multipoles.} If one approximates the discrete ellipticity catalog by distributing the $w_i$ and the $w_i \gammacs{,i}$ on a regular spatial grid with pixel size $\Delta$ and elements $w^{(\Delta)}_i$ and $(w\gammac)_i^{(\Delta)}$ then for any radial bin with lower bound $\Theta_{\mathrm{low}} \gg \Delta$ this approximation will give very similar results compared to the discrete catalog. The corresponding expression for the $G_n$ and $W_n$ can now be reformulated in terms of a convolution:
\begin{align}
    G_n^\Delta\left(\btheta_i^{(\Delta)};\Theta\right) &= \sum_{k=1}^{N_\mathrm{pix}^{\Delta}}  (w\gammac)_k^{(\Delta)} \ \ee^{\ii n \varphi_{ik}^{(\Delta)}} \ \binfunc\left(\theta_{ik}^{(\Delta)}\in\Theta \right)
    \nonumber \\ 
    &= \left[(w\gammac)^{(\Delta)} \star g_{n;\Theta}^{(\Delta)}\right]\left(\btheta_i^{(\Delta)}\right) \ ,
    \\
    W_n^\Delta\left(\btheta_i^{(\Delta)};\Theta\right) &=
    \left[w^{(\Delta)} \star g_{n;\Theta}^{(\Delta)}\right]\left(\btheta_i^{\left(\Delta\right)}\right) \ ,
\end{align}
where all quantities with the superscript $(\Delta)$ are evaluated at the pixel centers and ``$\star$'' denotes the convolution operator. The transformation to multipole components $\Upsilon^\times_{\mu,n}$ is equivalent to the discrete case \eqref{eq:Upsilon0_Derivation}, but the outer sum now runs over the number of pixels $\NpixDelta$ instead of the number of galaxies:
\begin{align}\label{eq:UpsilonmuGrid}
    \Upsilon_{0,n}^\times(&\Theta_1,\Theta_2) \nonumber \\ &= 
    -\sum_{i=1}^{\NpixDelta} \left(w\gammac\right)^{(\Delta)}_i\ 
    G_{n-3}^{(\Delta)}\left(\btheta_i^{(\Delta)};\Theta_1\right) \ G_{-n-3}^{(\Delta)}\left(\btheta_i^{(\Delta)};\Theta_2\right)
    \ .
\end{align}
The possibility of using FFT methods for computing the convolution gives a time complexity of the grid-based estimator of $\mathcal{O}\left(N_\mathrm{pix}^{(\Delta)} \log\left(N_\mathrm{pix}^{(\Delta)}\right)\right)$, significantly outperforming all of the previously mentioned estimators whenever applicable.
\subsection{Combined estimator}\label{ssec:CombinedEstimator}
For an ellipticity catalogue from a cosmic shear survey, we can now devise an efficient estimation procedure of the shear 3PCF as depicted in \figref{fig:EstimatorGeneralStrategy}. In the first step, we choose some scale $\theta_{\mathrm{disc}}$ up until which the multipoles are computed using the discrete prescription outlined in \secref{ssec:MultipoleDecomposition}. We then pick a base resolution scale $\Delta_1 \equiv \Delta$ and some coarser resolutions $\Delta_d \equiv 2^d\Delta \ (d=2,3,...)$ and distribute the ellipticity catalog on the corresponding grids. 
For each of the scales  $\theta>\theta_{\mathrm{disc}}$ we now select some resolution $\Delta_i(\theta)$ and compute the $G_n$ using the grid-based method from \secref{ssec:GridApproximation}. To allocate the multipoles of two scales for which the $G_n$ were computed with different methods, the $G_n$ with the higher resolution need to be mapped to the coarser grid before the $\Upsilon^\times_{\mu,n}$ can be updated. In particular, for the case where one of the $G_n$ is computed via the discrete method and the other one via the grid-based one, we first define the quantity $G_{n}'^{(\Delta)}\left(\btheta^{(\Delta)}_{p_{i}};\Theta_2\right)$ as the value of $G_n^{\Delta}$ at the pixel $p$ in which the $i$th galaxy resides, normalized by the number of galaxies in the $p$th pixel. We now have
\begin{align}\label{eq:CombinedEstimatorRegridding}
    \Upsilon^\times_{0,n}(&\Theta_1,\Theta_2) \nonumber \\
    &= 
    \sum_{i=1}^{N_{\mathrm{gal}}} w_i\gamma_{\mathrm{c},i} \ G_{n-3}^{\mathrm{disc}}(\btheta_i;\Theta_1) \ G_{-n-3}'^{(\Delta)}\left(\btheta^{(\Delta)}_{p_{i}};\Theta_2\right)
    \nonumber \\ &= 
    \sum_{p=1}^{\NpixDelta} G_{-n-3}^{(\Delta)}\left(\btheta^{(\Delta)}_{p};\Theta_2\right) \sum_{i=1}^{N_{\mathrm{gal},p}^{(\Delta)}} w_{i} \gamma_{\mathrm{c},i} \ G_{n-3}^{\mathrm{disc}}(\btheta_{i};\Theta_1)
    \nonumber \\ &\equiv
    \sum_{p=1}^{\NpixDelta} \left(w\gamma G_{n-3}^{\mathrm{disc}}\right)^{(\Delta)}\left(\btheta^{(\Delta)}_p;\Theta_1\right) \ G_{-n-3}^{(\Delta)}\left(\btheta^{(\Delta)}_p;\Theta_2\right) \ ,
\end{align}
where the second sum in the second line runs over all $N_{\mathrm{gal},p}^{(\Delta)}$ galaxies residing in the $p$th pixel. This implies that the products $w\gamma G_{n-3}^{\mathrm{disc}}$ need to be evaluated at the \textit{true} galaxy positions before they are mapped to the corresponding pixel. 
In \appref{app:EstimatorComplexity} we give more explicit details on the time and space complexity of the combined estimator.
\subsection{Edge corrections}\label{ssec:Norm3PCF}
Looking back at the general form of the estimator for the shear 3PCF \Eqbr{eq:3PCF_tripletestimator} we see that for a brute-force-like estimator the shear 3PCF $\Gamma_\mu$ can be evaluated by a simple division of the two correlators $\Upsilon_\mu$ and $\mathcal{N}$ for each angular- and polar bin combination. On the other hand, when working in the multipole basis we have
\begin{align}
    \Gamma^\times_\mu(\Theta_1, \Theta_2,\phi) = \frac{1}{2\pi} \sum_{n=-\infty}^\infty \Gamma^\times_{\mu,n}(\Theta_1,\Theta_2) \ \ee^{\ii n\phi} \ .
\end{align}
To work out the analog of the division in the multipole basis we first multiply \Eq{eq:3PCF_tripletestimator} by $\mathcal{N}$ and then insert the multipole expansion of all quantities:
\begin{align}\label{eq:Edgcorr_helper1}
    \sum_{n,n'=-\infty}^\infty \ \Gammabarx{\mu,n}{\Theta_1,\Theta_2}
    \ &
    \mathcal{N}_{n'}(\Theta_1,\Theta_2) \ \ee^{\ii(n+n')\phi}
    \nonumber \\
    &=
    \sum_{n=-\infty}^\infty
    \Upsilon_{\mu,n}^\times(\Theta_1,\Theta_2) \ \ee^{\ii n \phi} \ .
\end{align}
Multiplying \Eq{eq:Edgcorr_helper1} by $\ee^{-\ii\ell\phi}$ and integrating over $\phi$ we find:
\begin{align}\label{eq:ModeCoupling}
    \sum_{n=-\infty}^\infty \mathcal{N}_{\ell-n}(\Theta_1,\Theta_2) \ \Gammabarx{\mu,n}{\Theta_1,\Theta_2} = \Upsilon_{\mu,\ell}^\times(\Theta_1,\Theta_2) \ .
\end{align}
Thus, we see that in general there exists a coupling between different multipoles. While \Eq{eq:ModeCoupling} formally describes an infinite system of equations, one expects that $\mathcal{N}_{\ell-n}$ is dominated by its diagonal, such that truncating the expansion at some $n_{\mathrm{max}}$ will provide a sufficient approximation.\footnote{If we assume the 3PCF to not rapidly vary over small angular scales, we expect the amplitude of the $\mathcal{N}_n$ to approach zero for large multipoles $n$. Choosing a sufficiently large $n_{\mathrm{max}}$ then implies that the quantity $\mathcal{N}_{\ell-n}$ will have its largest contributions for $n\approx \ell$.}
Inverting the resulting $(2n_{\mathrm{max}}+1)$ - dimensional system equips us with an expression for the edge-corrected multipoles of the shear $3$PCF:
\begin{align}\label{eq:GammaMultipolesWithEC}
\Gammabarx{\mu,n}{\Theta_1,\Theta_2} = 
\sum_{\ell=-n_{\mathrm{max}}}^{n_{\mathrm{max}}}{}\left(\mathcal{C}^{-1}\right)_{n\ell}(\Theta_1,\Theta_2) \ \frac{\Upsilon^{\times}_{\mu,\ell}(\Theta_1,\Theta_2)}{\mathcal{N}_{0}(\Theta_1,\Theta_2)} \ ,
\end{align}
where we defined the components of the mode coupling matrix $\mathcal{C}$ as 
\begin{align}
    \mathcal{C}_{\ell n}(\Theta_1,\Theta_2) =  \frac{\mathcal{N}_{\ell-n}(\Theta_1,\Theta_2)}{\mathcal{N}_{0}(\Theta_1,\Theta_2)} \ .
\end{align}
For a uniform distribution of tracers, $\mathcal{N}$ does not carry any angular dependence such that $\mathcal{C}$ becomes the identity matrix. We note that in the cosmological context, a nonuniform distribution of tracers can either be realized by a clustering of the tracers or a nontrivial survey geometry. We further note that this mode coupling effect has already been explored in \cite{SlepianEisenstein2016} in the context of edge-correcting the 3PCF of scalar fields.
\subsection{Including tomography}\label{ssec:TomoEstimator}
Our estimator can straightforwardly be generalized to a tomographic setting with $n_z$ tomographic redshift bins $ \{Z_k\} \ (k\in \{1,\cdots,n_z\})$, where we define the galaxy at the location $\mathbf{X}_i$ in each galaxy triplet to lie in the $Z_i$th redshift bin. In the case of gridded data, we simply need to evaluate the $G_n^{\Delta}$ for each tomographic bin and then allocate the $\Upsilon_{\mu,n}^\times$ for all relevant combinations:
\begin{align}
&\Upsilon_{0,n}^\times(\Theta_1,\Theta_2;Z_1,Z_2,Z_3) = -\sum_{i=1}^{\NpixDelta} \left(w_i\gamma_{\mathrm{c},i}\right)^{(Z_1,\Delta)} 
\nonumber \\
    &\hspace{1.3cm}\times \  
    G_{n-3}^{(Z_2,\Delta)}(\btheta_i;\Theta_1) \ G_{-n-3}^{(Z_3,\Delta)}(\btheta_i;\Theta_2) \ , \\ 
    &G_{n}^{(Z_2,\Delta)}(\btheta_i;\Theta_1) = \left[(w\gammac)^{(Z_2,\Delta)} \star g_{n;\Theta}^{(\Delta)}\right](\btheta_i) \ .
\end{align}
For discrete data the modifications are similar, but one additionally needs to keep track of the redshift bins of both galaxies contributing to  the $G_n$:
\begin{align}\label{eq:tomoGn} 
\Upsilon_{0,n}^\times(\Theta_1,\Theta_2;Z_1,Z_2,Z_3) 
&= - \sum_{i=1}^{\Ngal^{(Z_1)}}\left(w_i\gamma_{\mathrm{c},i}\right)^{(Z_1,\Delta)} \nonumber \\
\times \ G_{n-3}^{(Z_2,\mathrm{disc})}&(\btheta_i;\Theta_1) \ G_{-n-3}^{(Z_3,\mathrm{disc})}(\btheta_i;\Theta_2) \ , \\ G_{n}^{(Z_2,\mathrm{disc})}(\btheta_i;\Theta) &\equiv 
\sum_{k=1}^{\Ngal^{(Z_2)}} w_k\gammacs{,k}  \ g_{n;\Theta}(\btheta_{ik}) \ .
\end{align}
The transformation to the natural components works analogously to \Eq{eq:GammaMultipolesWithEC}.

\begin{figure*}
  \includegraphics[width=.49\linewidth]
  {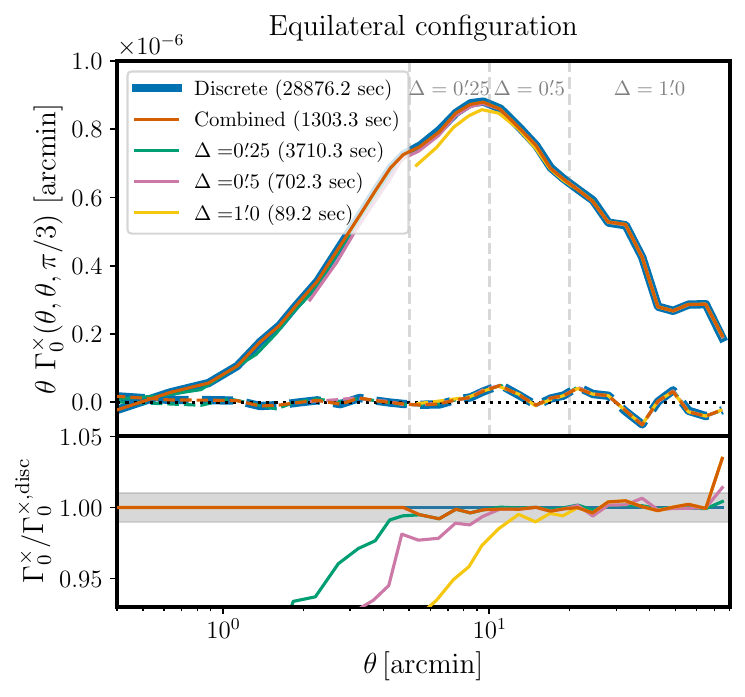} 
  \includegraphics[width=.49\linewidth]
{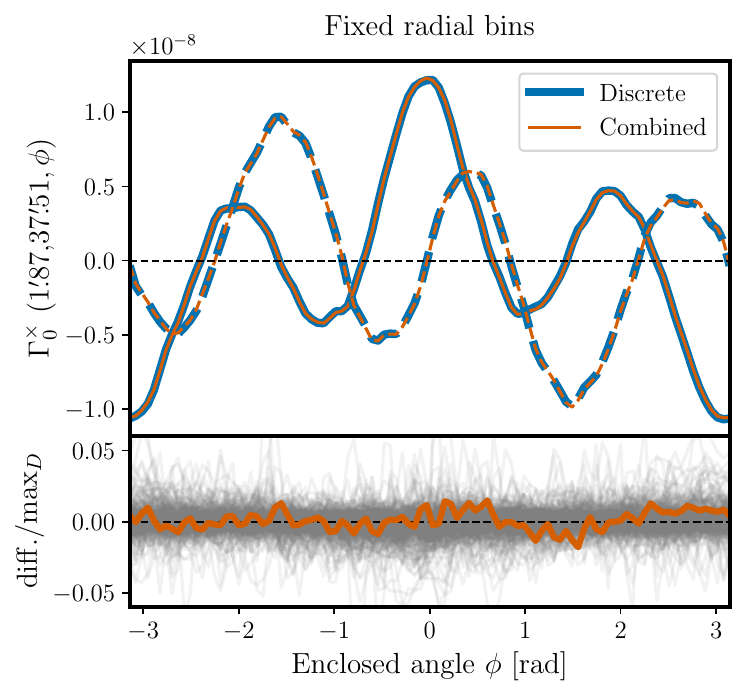}
\caption{Performance of the combined estimator on a noiseless mock catalog with $\approx 3.1\times 10^6$ ellipticities on a $100 \, \deg^2$ area. In the upper panels, solid lines indicate the real part of the 3PCF, while dashed lines represent the imaginary part. \textit{Left hand side}: Comparison of the discrete, various grid-based, and a combined implementation of the multipole estimator for equilateral configurations of the zeroth natural component of the shear 3PCF. The timings correspond to the runtime on $8$ CPU cores and the dashed vertical lines indicate the regions of constant grid resolution for the combined estimator. In the bottom panel, we show the ratios of the real parts between all approximate implementations and the discrete estimator. The grey-shaded region displays a one-per-cent interval. We note that by construction the curve corresponding to the combined estimator always lies on top of the curve corresponding to the resolution $\Delta_d$ in the interval $20\Delta_d \leq \Theta_{\mathrm{low}} \leq \Theta_{\mathrm{up}} \leq 40\Delta_d$.
\textit{Right-hand side}: Comparison of the discrete and the combined estimator for different triangle shapes. We fix two triangle sides and vary the enclosing angle $\phi$. In the bottom panel, the thick red line shows the difference between the real part of both estimators, normalized by the largest value of the discrete estimator for the given radial bin combination $\mathrm{max}_D$. The grey lines display the corresponding ratios for all other radial bin combinations, where again the normalization is done with respect to the largest value of the discrete estimator for the corresponding radial bin combination.
}
\label{fig:MultiScaleValidation}
\end{figure*}
%
\section{Validation of the estimator}\label{sec:Validation}
We validate our estimator on the Scinet-LIghtCone Simulations (SLICS) simulation suite \citep{HarnoisDerapsetal2018}. They consist of a suite of dark matter-only N-body simulations which track the evolution of $1536^3$ particles within a box of sidelength $505\, h^{-1}$Mpc. From each simulation convergence maps are constructed up until $z=3$ within a field of view of $100 \, \deg^2$ on a grid of $7745^2$ pixels. For this work, we use the KiDS-450-like ensemble in which $819$ fully independent ellipticity catalogs mimicking the redshift distribution and number density of the KiDS-450 data \citep{DeJongetal2017} are constructed from the past light-cones. We further use a maximum multipole of $n_{\mathrm{max}}\equiv 20$ for all tests presented in this and in the following section. For the remainder of this work, when transforming the shear 3PCF from its multipole basis to the real space basis, we use 100 linearly-spaced angular bins between $[0,2\pi]$, independent of the chosen value for $n_{\mathrm{max}}$.
\subsection{Accuracy of the combined estimator}
For testing the applicability of the grid-based approximation and the associated combined estimator we apply all of those schemes to a single noiseless mock catalog of the SLICS ensemble. In particular, we evaluate the 3PCF multipoles in $33$ log-spaced radial bins between $0\farcm25$ and $80'$ which are then transformed to the real-space shear 3PCF. In \figref{fig:MultiScaleValidation} we compare the runs using the discrete estimator, the grid-based approximation for resolution scales $\Delta \in \{0\farcm25,0\farcm5,1\farcm0\}$, as well as the combined estimator that uses the grid-based approximation for scales $20\Delta \leq \Theta_{\mathrm{low}} \leq \Theta_{\mathrm{up}} \leq 40\Delta$ and afterwards switches to the next higher resolution scale up until $\Delta=1'$.\footnote{As we define the edges of the angular bins to always contain $\{20\Delta_d, 40\Delta_d\}$ for all resolution scales, we do not need to consider edge cases like $ \Theta_{\mathrm{low}} \leq 40\Delta \Theta_{\mathrm{up}}$. We note that while these bounds are not hard-coded, we will use them for the remainder of this work whenever making use of the combined estimator.} For this binning scheme, the combined estimator performed $\approx 20$ times faster than the discrete estimator, even outperforming the pure grid-based approximation of $\Delta=0\farcm25$. Focusing our attention on the equilateral configurations we see that the grid-based approximation is in good agreement with the discrete estimator once the separation $\Theta$ is much larger than the pixelization scale $\Delta$, but it starts to bias the statistics once the separation is less than around ten times the pixelization scale. For the equilateral configurations, the combined estimator is, by construction, equal to the discrete estimator for $\Theta<5'$ and then follows the measurements of the grid-based implementations. For our example shown in \figref{fig:MultiScaleValidation}, the combined estimator is in agreement with the discrete estimator at the sub-per-cent level.

For assessing the accuracy of the blocks in the $(\Theta_1,\Theta_2)$-plane shown in \figref{fig:EstimatorGeneralStrategy}, in which the first scale is evaluated using the discrete prescription while the second scale uses the grid-based approximation, we pick one such combination of radial bins and then vary the enclosed angle $\phi$ between them. For the example shown in the right-hand panel in \figref{fig:MultiScaleValidation} we choose a scale for which the $G_n^{(\Delta)}$ are evaluated on a grid with $\Delta=1'$ and we find an excellent agreement between both methods. We also show the ratios between the two schemes for all other bin combinations. Although for the majority of the curves, the ratio is very close to unity, there do appear some spikes for which the relative difference between both estimators is at the five percent level. However, those bin configurations correspond to cases in which are only a few triplets present and therefore they will not carry significant signal-to-noise in realistic data with non-vanishing shape noise.
\subsection{Comparison with \treecorr}
\begin{figure}
  \centering
  \includegraphics[width=.99\linewidth]{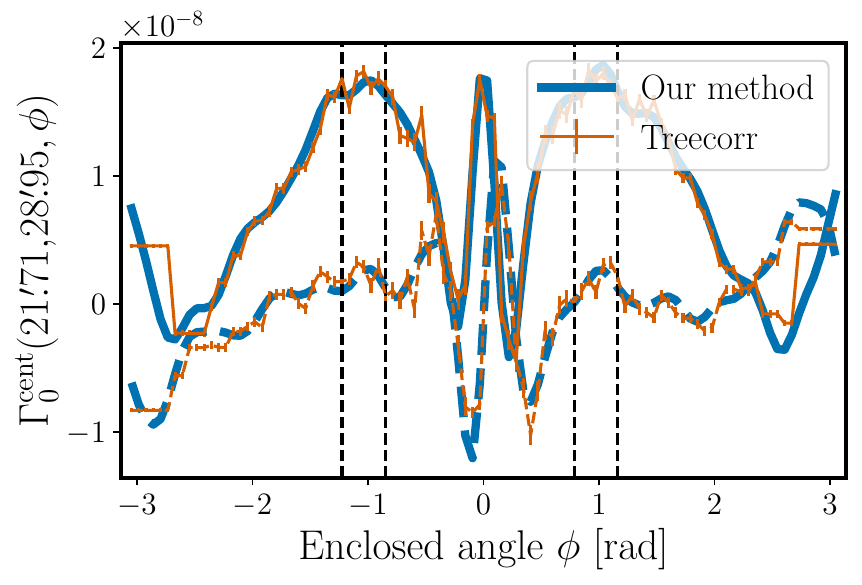}
  \includegraphics[width=.99\linewidth]{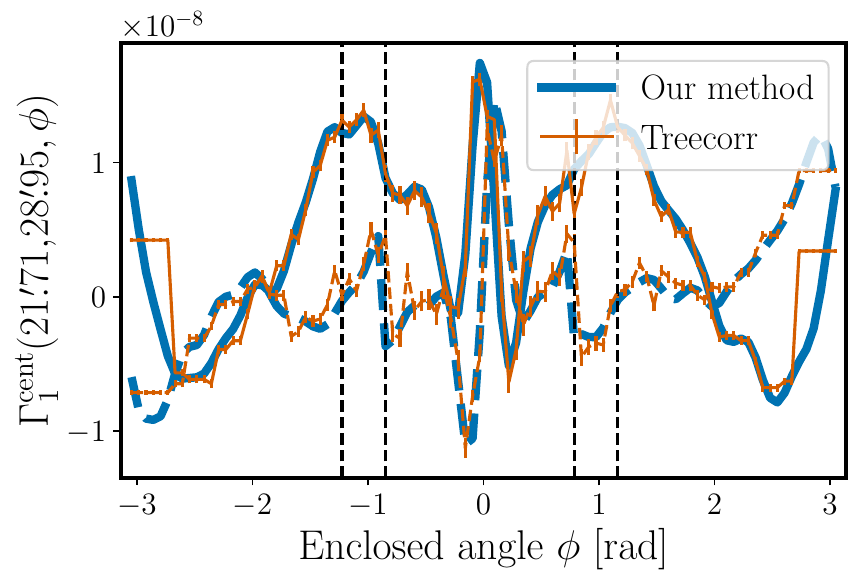}
\caption{The first two natural components of the shear 3PCF on a catalog with $\approx 3.1 \times 10^5$ shapes on a $100 \, \mathrm{deg}^2$ area. We fix two triangle sides and vary the enclosing angle $\phi$. Solid lines indicate the real part
of the 3PCF while dashed lines stand for the imaginary part. The regions between the dashed black lines correspond to the intervals in which the ordering of the lengths of the triangle is constant. We note that due to the different binning parametrizations of \treecorr and our estimator, we do not expect perfect agreement between the curves.}
\label{fig:GammaTreeCorrCompare}
\end{figure}
The state-of-the-art method for computing shear correlation functions is based on tree codes -- the most well-known implementation is the publicly available code \treecorr \ \citep{Jarvisetal2004}\footnote{For the tests presented in this subsection we are using version 4.2.9.}. In there, the $\overline{\Gamma}_\mu$ are estimated by explicitly accounting for all triplets, where for an efficient evaluation, \treecorr first organizes the data in a hierarchical ball-tree structure from which the 3PCF is then computed. A further speedup can be achieved when allowing the bin edges to be fuzzy; while for reasonable values of the associated \texttt{bin\_slop} parameter, this approximation does not bias the mean, but it increases the measurement uncertainty \citep{Seccoetal2022}. For an efficient binning of triangle shapes \treecorr uses the quantities $(r, u, v)$ that are related to the three triangle sidelengths $d_1 \geq d_2 \geq d_3$ as
\begin{align}
    r = d_2 \ \ \ ; \ \ \ 
    u = \frac{d_3}{d_2} \ \ \ ; \ \ \
    v = \pm \frac{d_1-d_2}{d_3} \ ,
\end{align}
such that with $u \in [0,1]$, $v \in [-1,1]$ all triangle configurations with $d_1 \geq d_2 \geq d_3$ are covered; as shown in \cite{Jarvisetal2004} this coverage is sufficient to obtain integrated measures of the 3PCF, such as the third-order aperture mass. 

In \figref{fig:GammaTreeCorrCompare} we compare the estimated components of the first two natural components $\Gamma^{\mathrm{cent}}_{0,1}$ on a single noiseless mock catalog of the SLICS ensemble. In order to facilitate an accurate matching between the two binning parametrizations one needs to use very fine bins for both estimators. For our test, we compute the 3PCF between $5'$ and $50'$, where for the multipole-based estimator we use $40$ logarithmically-spaced radial bins without making use of the grid-based approximation. For the \treecorr measurement we choose the same $r$-binning and use $20 \, (40)$ linearly spaced bins covering the maximum range of the $u \, (v)$ parameters. To achieve a reasonable runtime we only select every tenth galaxy, leaving a total of $\approx 3.1 \times 10^5$ ellipticities. The associated measurements were carried out on $16$ CPU cores and took $\approx 1.5$ minutes for the discrete multipole-based estimator and $\approx 164$ minutes for \treecorr when using a value of $0.8$ for the \texttt{bin\_slop} parameter. In \figref{fig:GammaTreeCorrCompare} we find overall good agreement between both estimators across the different triangle configurations; the only significant differences are visible for strongly squeezed and folded triangles. 

Besides the significantly lower computational cost compared to the tree code, the multipole-based estimator also automatically allows for a fine $\phi$-binning of the shear 3PCF and thus provides a good approximation even for strongly degenerate triangles.\footnote{We recall that while the angular binning can be made arbitrarily small, the angular scale of the smallest features in the estimated 3PCF depends on the largest considered multipole $n_{\mathrm{max}}$.} Finally, we note that the relative speedup with respect to \treecorr increases when the combined estimator is employed and when the number density of galaxies increases. For a thorough scaling analysis of the combined estimator, we again refer to \appref{app:EstimatorComplexity}.
\section{Application to third-order aperture mass measures}\label{sec:ApertureMass}
Due to the large size of a data vector built from an estimate of the shear 3PCF, it is unfeasible to perform a cosmological analysis using this probe. The preferred statistic is the third-order aperture mass, which provides both a compression of the data vector and a separation of the third-order shear signal in its $E$- and $B$-modes.
\subsection{Aperture mass} 
\label{ssec:ApertureMass}
The aperture mass introduced in \cite{Kaiser1995,Schneider1996} is a measure for the projected overdensity within a circular region of radius $\theta$ around some location $\bvartheta$;
\begin{align}\label{eq:MapDefinition}
    \map(\bvartheta;\theta) = \int \dd^2 \bvartheta' \ U_\theta\left(|\bvartheta'|\right) \ \kappa(\bvartheta+\bvartheta') \ ,
\end{align}
where the filter function $U_\theta$ is compensated, meaning that aperture mass measures evade the mass-sheet degeneracy \citep{Falcoetal1985}. Within the flat-sky approximation, the aperture mass can also be written as the real part of a complex aperture measure being defined in terms of the shear
\begin{align}\label{eq:MapGamma}
   \mathcal{M}(\bvartheta;\theta) 
    &=
   \map(\bvartheta;\theta) + \ii \mperp(\bvartheta;\theta) \nonumber \\
    &=
    \int \dd^2 \bvartheta' \ Q_\theta\left(|\bvartheta'|\right) \ \gamma(\bvartheta+\bvartheta';\varphi') \ ,
\end{align}
in which $Q_\theta$ is another filter function uniquely related to the choice of the $U_\theta$ filter, and $\varphi'$ denotes the polar angle of the separation vector $\bvartheta'$. With the aperture measure being able to separate the shear signal into its $E$- and $B$-modes, it can be used for both gaining cosmological information and pinning down potential systematics. For this work, we use the exponential filter introduced in \cite{Crittendenetal2002},
\begin{align}
    \label{eq:CrittendenQ}
    Q_\theta(\vartheta) = \frac{\vartheta^2}{4\pi\theta^4}\exp[\left(-\frac{\vartheta^2}{2\theta^2}\right)] \ .
\end{align}

\subsection{Third-order aperture mass measures}
The $E$/$B$ decomposition also extends to the third-order aperture mass measures. For example, combining three aperture measures as 
\begin{align}
    \mapthreeens (\theta_1,\theta_2,\theta_3)
    \equiv
    \left\langle
    \mathcal{M}_\mathrm{ap}(\theta_1) \ 
    \mathcal{M}_\mathrm{ap}(\theta_2) \ 
    \mathcal{M}_\mathrm{ap}(\theta_3)
    \right\rangle
\end{align}
one can see from \Eq{eq:MapDefinition} that the resulting quantity depends on the third-order correlation function of the convergence field. By writing down the non-redundant products that appear when combining three complex measures $\mathcal{M}$,
one obtains four real-valued quantities $\mapthreeens, \left\langle\mathcal{M}_\mathrm{ap}^2\mathcal{M}_{\times}\right\rangle$, $\left\langle\mathcal{M}_\mathrm{ap}\mathcal{M}_{\times}^2\right\rangle$ and $\left\langle\mathcal{M}_{\times}^3\right\rangle$, that are all related to a filtered version of the third-order shear correlator. Of those, only $\mapthreeens$ carries cosmological information, while the two measures $\left\langle\mathcal{M}_\mathrm{ap}^2\mathcal{M}_{\times}\right\rangle$ and $\left\langle\mathcal{M}_{\times}^3\right\rangle$ vanish for parity symmetric shear fields \citep{Schneider2003}, and a statistically significant signal of $\left\langle\mathcal{M}_\mathrm{ap}\mathcal{M}_{\times}^2\right\rangle$ indicates the presence of $B$-modes \citep[][hereafter SKL05]{Schneideretal2005}. Using the centroid projection \Eq{eq:centroidprojection} for the natural components \citetalias{Schneideretal2005} derived an explicit form of the transformation between the $\Gamma_\mu^\mathrm{cen}$ and the third-order aperture mass measures that are defined in terms of two third-order correlators of the complex aperture measure $\mathcal{M}$:
\begin{align}\label{eq:MFrom3pcfTheory}
    \expval{\mathcal{M}^3}(\theta_1,\theta_2,\theta_3) &= \int_0^\infty \dd \vartheta_1 \int_0^\infty \dd \vartheta_2 \int_0^{2\pi}\dd \phi 
    \nonumber\\ &\hspace*{-1cm} \times \Gamma_{0}^\mathrm{cen}(\vartheta_1,\vartheta_2,\phi) \ F_0(\theta_1,\theta_2, \theta_3;\vartheta_1,\vartheta_2,\phi)\ , \nonumber\\
     \expval{\mathcal{M}^*\mathcal{M}^2}(\theta_1,\theta_2,\theta_3) &= \int_0^\infty \dd \vartheta_1 \int_0^\infty \dd \vartheta_2 \int_0^{2\pi}\dd \phi
    \nonumber \\ &\hspace*{-1cm} \times \Gamma_{1}^\mathrm{cen}(\vartheta_1,\vartheta_2,\phi) F_1(\theta_1,\theta_2,\theta_3;\vartheta_1,\vartheta_2,\phi) \ ,
\end{align}
where the filter functions $F_{0,1}$ in our convention only differ from the ones in \citetalias{Schneideretal2005} by cyclic permutations of the indices.\footnote{The origin of the permutation relative to the equations in \citetalias{Schneideretal2005} lies in the different conventions of the defining correlators of the $\Gamma_\mu$ in \Eq{eq:Gamma0_def}. This only affects the $F_1$ filter, as the $F_0$ filter is symmetric under permutations of the $\mathbf{q}$-vectors.} We give the explicit expressions of the filters in our convention in \appref{app:3PCFTrafo}.

While these two correlators are sufficient for the theoretical conversion, this might not be the case when the shear 3PCF is estimated from noisy data which is distributed over several tomographic redshift bins. In this case, one needs to include two additional correlators if the full signal-to-noise should be retrieved:
\begin{align}\label{eq:MFrom3pcfTheoryb}
    \expval{\mathcal{M} \mathcal{M}^* \mathcal{M}}(\theta_1,\theta_2,\theta_3) &=\int_0^\infty \dd \vartheta_1 \int_0^\infty \dd \vartheta_2 \int_0^{2\pi}\dd \phi
    \nonumber \\ &\hspace*{-1cm} \times \Gamma_{2}^\mathrm{cen}(\vartheta_1,\vartheta_2,\phi) F_2(\theta_1,\theta_2,\theta_3;\vartheta_1,\vartheta_2,\phi) \ , \nonumber\\
    \expval{\mathcal{M}^2\mathcal{M}^*}(\theta_1,\theta_2,\theta_3) &= \int_0^\infty \dd \vartheta_1 \int_0^\infty \dd \vartheta_2 \int_0^{2\pi}\dd \phi
    \nonumber \\ &\hspace*{-1cm} \times \Gamma_{3}^\mathrm{cen}(\vartheta_1,\vartheta_2,\phi) \ F_3(\theta_1,\theta_2,\theta_3;\vartheta_1,\vartheta_2,\phi) \ .
\end{align}
 The individual third-order aperture mass measures then follow by a linear combination of the previous quantities; in particular, for $\mapthreeens$, one has:
\begin{align}\label{eq:Mapthreefrom3PCF}
&\mapthreeens(\theta_1,\theta_2,\theta_3) \nonumber \\
&\quad \quad =\Re\Large[\expval{\mathcal{M}^2 \mathcal{M}^*}(\theta_1,\theta_2,\theta_3)+\expval{\mathcal{M} \mathcal{M}^* \mathcal{M}}(\theta_1,\theta_2,\theta_3) \nonumber \\
& \quad \quad \quad +\expval{\mathcal{M}^*\mathcal{M}^2}(\theta_1,\theta_2,\theta_3)+\expval{\mathcal{M}^3}(\theta_1,\theta_2,\theta_3)\Large]/4 \ ,
\end{align}
and the linear combinations of the remaining aperture measures can be constructed in a similar fashion.

For a tomographic setup, the structure of the equations above remains equal, but we make the substitution 
\begin{align}
    \Gamma_\mu^{\mathrm{cent}}(\vartheta_1,\vartheta_2, \phi) \rightarrow \Gamma_\mu^{\mathrm{cent}}(\vartheta_1,\vartheta_2, \phi; Z_1, Z_2, Z_3) \ ,
\end{align}
where the $\Gamma_\mu^{\mathrm{cent}}(\vartheta_1,\vartheta_2, \phi; Z_1, Z_2, Z_3)$ are estimated as described in \secref{ssec:TomoEstimator}. This in turn means that all the other quantities, like $\expval{\mathcal{M}^3}$ or $\mapthreeens$, also become functions of the tomographic redshift bin combination $(Z_1, Z_2, Z_3)$. We note that while the third-order aperture mass is symmetric with respect to permutations of the $Z_i$, this is not the case for the shear 3PCF and therefore one needs to average over all possible $Z_i$ permutations to obtain the full signal-to-noise. We again refer to \appref{app:3PCFTrafo} for all explicit expressions.
\subsection{Estimators of third-order aperture measures}
\subsubsection{Estimation via shear 3PCF}
As established in the previous subsection one can estimate the third-order aperture mass measures by integrating over the 3PCF in the centroid basis. For doing this in practice we first use the multipole-based estimator \Eqbr{eq:GammaMultipolesWithEC} introduced in \secref{sec:Estimator} to obtain the shear 3PCF in the $\times$-projection, then transform them to the centroid basis via Eqs. (\ref{eq:X2Centroid0}-\ref{eq:X2Centroid3}), and finally perform the numerical integrals contained in \Eq{eq:Mapthreefrom3PCF}. In \appref{app:Map3Convergence} we show the level of accuracy to which the 3PCF needs to be computed in order to guarantee a stable integration. As the computational bottleneck of this pipeline is the estimation of the shear 3PCF, the same computational speedup applies to the estimation of the third-order aperture measures when using the multipole-based estimator instead of traditional estimation methods of the shear 3PCF.
\subsubsection{Estimation via the direct estimator}
The original estimator for the third-order aperture mass statistics in simulated data on an area $A$ has been introduced in \cite{Schneideretal1998}. Instead of evaluating the statistics in terms of the shear 3PCF, it directly samples third-order generalization of \Eq{eq:MapGamma},
\begin{align}
    \left\langle\mathcal{M}_{\mathrm{ap}}^3\right\rangle (\theta_1,\theta_2,\theta_3)
    &=
     \int \frac{\dd^2 \bvartheta}{A}   \nonumber \\
    \times \ \prod_{i=1}^3 &\int \dd^2 \bvartheta_i \ Q_{\theta_i}\left(|\bvartheta_i|\right) \ \gammat(\bvartheta+\bvartheta_i;\phi_i) \ ,
\end{align} 
from the ellipticities at the galaxy positions; i.e. at some angular position $\bvartheta$ one has
\begin{align}\label{eq:MapDirectEstimator}
    \mapthreehat(&\bvartheta; \theta_1,\theta_2,\theta_3)  \nonumber \\
    &=\frac{\sum_{i\neq j \neq k}^{N_{\mathrm{gal}}} 
    w_i w_j w_k \
    Q_{\theta_1,i} Q_{\theta_2,j}Q_{\theta_3,k} \ 
    \gamma_{\mathrm{t},i} 
    \gamma_{\mathrm{t},j} 
    \gamma_{\mathrm{t},k}}
    {\sum_{i\neq j \neq k}^{N_{\mathrm{gal}}}
    w_i w_j w_k \
    Q_{\theta_1,i} Q_{\theta_2,j}Q_{\theta_3,k} \ } \ ,
\end{align}
where we abbreviated $Q_{\theta_l,i}\equiv Q_{\theta_l}(|\bvartheta_i-\bvartheta|)$ and $\gamma_{\mathrm{t},i} \equiv \gammat (\bvartheta_i; \zeta_i)$ with $\bvartheta_i$ being the angular position of the $i$th galaxy and $\zeta_i$ denoting the direction $\bvartheta_i-\bvartheta$. The corresponding estimate for $\mapthreeens$ is then obtained by covering the survey footprint with apertures and averaging over the individual estimates. As shown in \cite{Porthetal2021}, the nested sums appearing in \Eq{eq:MapDirectEstimator} can be decomposed such that the estimation procedure scales linearly in the number of galaxies in the survey\footnote{We note that, as the direct estimator assumes the apertures to be fully contained within the survey footprint, its application to a nontrivial survey geometry is not easily possible.}. Although the employed $Q$-filter \eqref{eq:CrittendenQ} does formally have infinite support, one can show that by only including galaxies separated by at most $4\theta$ from the aperture center, one recovers a practically unbiased result \citep{Heydenreichetal2022}. Thus, in order to avoid border effects we do not sample apertures in regions which are separated by less than $4\max(\theta_1,\theta_2,\theta_3)$ from the fields' boundary. We recall that while the direct estimator can be used on unmasked simulated data, like the SLICS ensemble, this is not the case for real data with a nontrivial angular survey mask.

\subsection{Comparison of both estimators}
\begin{figure}
  \centering
  \includegraphics[width=.9\linewidth]{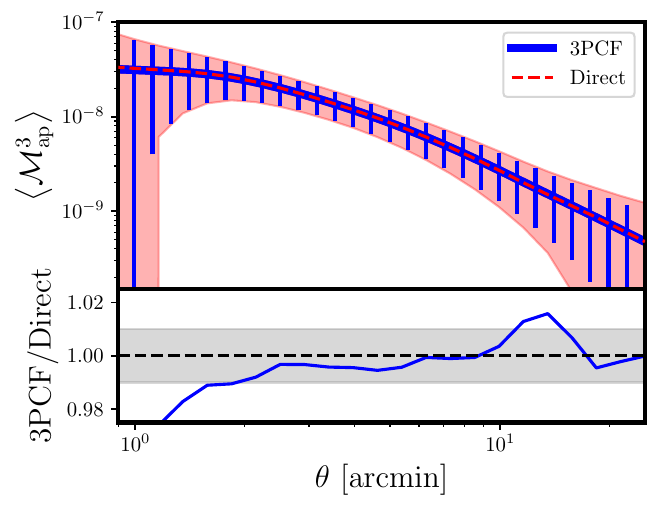}
\caption{Comparison of the third-order aperture mass statistics between the multipole-based estimator (blue line with error bars) and the direct estimator (red dashed line with error band) in the SLICS ensemble. In the lower panel, we plot the ratio between the two measurements (blue line) and show a one percent interval as the grey-shaded region.}
\label{fig:Map3DirectCompare}
\end{figure}

We compare the results for both estimation procedures on the SLICS ensemble, using all 819 mock catalogs. We first estimate the 3PCF in the interval $[0\farcm3125, 160']$ using a logarithmic bin width of $b\equiv \frac{\Theta_{\mathrm{up,i+1}}}{\Theta_{\mathrm{up,i}}} = 1.05$. We then transform the 3PCF to the equal-scale aperture statistics, where for the aperture radii we choose $30$ logarithmically spaced values between $[1',25']$. Finally, we estimate $\mapthreeens$ using the direct estimator \eqref{eq:MapDirectEstimator} for the same set of aperture radii. 
In \figref{fig:Map3DirectCompare}, we show the measurements obtained from both estimators, together with their standard deviation across the SLICS ensemble. We find a percent level agreement on aperture scales between $2'$ and $10'$; while for smaller aperture radii, the multipole-based estimator begins to yield systematically lower results, for larger aperture radii, there appears to be a scatter around the zero axis. Both of these cases are expected. On the one hand, as the shear 3PCF has not been estimated down to zero separation, the integral transformation \eqref{eq:MFrom3pcfTheory} cannot be fully sampled for small aperture radii. \cite{Shietal2014} have demonstrated that this results in a systematic underprediction of the estimated $\mapthreeens$ for aperture radii of less than $\sim 5$ times the smallest angular separation considered in the 3PCF. For the large scales, on the other hand, one would not expect both estimators to perfectly agree with each other. Cutting off the survey boundary for the direct estimator implies that both estimators do not use the data in the same way. This can also be seen in the standard deviation of both estimators across the ensemble. While for small aperture radii, the error bars are of a similar size, the multipole-based estimator gradually extracts more signal-to-noise compared to the direct estimator for increasing aperture radii, as not all triangles can be accounted for in the latter due to the restricted sampling domain of the apertures. 
\section{Application to the KiDS-1000 data}\label{sec:KiDSMeasurement}
\subsection{The KiDS-1000 data}
As a first application of our estimator to real data we use the gold sample of weak lensing and photometric redshift measurements from the fourth data release of the Kilo-Degree Survey \citep{Kuijkenetal2019,Wrightetal2020,Hildebrandtetal2021,Giblinetal2021}.\footnote{The KiDS data products are public and available through \url{https://kids.strw.leidenuniv.nl/DR4/}} The gold sample consists of around $21$ million galaxies distributed over an area of $1006$ square degrees ($777$ square degrees after masking) and is further divided into five tomographic redshift bins.
\subsection{Covariance matrix}
\begin{figure*}
  \centering
  \includegraphics[width=.99\textwidth]{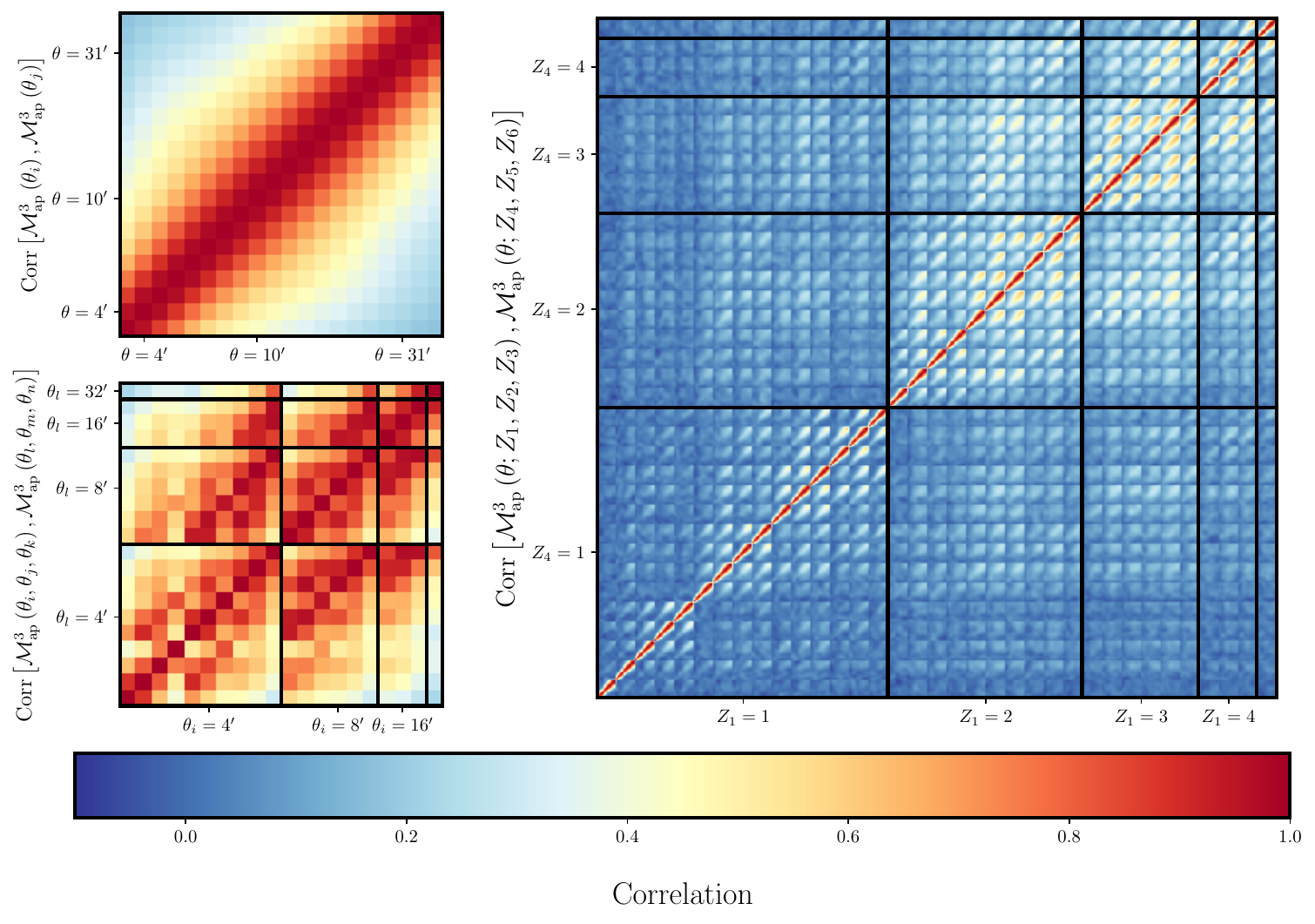}
\caption{Correlation matrices of third-order aperture mass measures in the T17 ensemble. Shown are the results for the non-tomographic, equal-scale statistics (top left), the non-tomographic unequal-scale statistics (bottom left), as well as the tomographic equal-scale statistics (right-hand side). We order the statistics such that $\theta_i \leq\theta_j \leq \theta_k$ and $Z_1 \leq Z_2 \leq Z_3$, meaning that each of the $35^2$ `small' squares in the covariance matrix on the right-hand side corresponds to the cross-covariance of the statistics for fixed tomographic redshift bin combinations.} \label{fig:TakahashiCorrcoef}
\end{figure*}
We obtain an estimate for the covariance matrix by using a suite of full-sky
gravitational lensing simulations introduced in \cite{Takahashietal2017}, hereafter the T17 ensemble. The underlying dark matter only $N$-body simulations are run on a $\Lambda$CDM cosmology with $\Omega_\mathrm{m}=0.279$,  $\Omega_{\Lambda}=1-\Omega_{\mathrm{m}}$, $h=0.7$, $\sigma_8=0.82$ and spectral index $n_\mathrm{s}=0.97$. We construct $1944$ mock ellipticity catalogs from the T17 ensemble. In particular, we match the angular positions, as well as the weights, the shape noise level per tomographic redshift bin, and the correlation between the measured ellipticities and the weights to the public KiDS-1000 ellipticity catalog \citep{Giblinetal2021}. We note that those specifications give rise to a different correlation structure as would be given by the SLICS ensemble, in which the galaxies are randomly placed within an unmasked area and in which each ellipticity carries an equal weight. For further details about the creation of the mocks, we refer to \cite{Burgeretal2023}. 

To measure the shear 3PCF via the multipole estimator we first decompose the survey area into $16$ smaller patches. We proceed along the lines of \appref{app:EstimatorOnSphere} to measure the 3PCF multipoles in each patch and then join them after which we obtain an unbiased estimate for the third-order aperture statistics. We measure the shear 3PCF in 38 angular bins in the interval $[0\farcm75,240']$, yielding a mean logarithmic angular bin width of $b\approx 1.15$. The choice of the lower bound is due to the pixelization scale of underlying shear maps in the T17 ensemble, which are given in a Healpix format with \texttt{nside}=8192, for which the individual pixels have a side length of $\approx 0\farcm4$. As we show in \appref{app:Map3Convergence}, this range of angular scales yields a numerically stable transformation of the shear 3PCF to the aperture mass measures for aperture scales between $4'$ and $40'$. We do not expect that much additional information is contained in larger aperture scales, as the third-order aperture mass only filters out non-Gaussian contributions of the cosmic shear field. As we aim to obtain an accurate covariance matrix for the unequal-scale aperture statistics in the non-tomographic setup we compute the multipoles up to $30$th order in this case. For the tomographic case we only consider equal aperture scales in the measurement, for which we show in \appref{app:Map3Convergence} that choosing $n_{\mathrm{max}}=10$ is sufficient. Finally, we transform the shear 3PCF to the aperture statistics on two sets of radii, the `fine-binning set' consisting of $30$ logarithmically-spaced radii in the interval $[4', 40']$ and the `coarse-binning set' for which $\theta \in \{4', 6', 8', 10', 14', 18', 22', 26', 32', 36'\}$. Using these configurations the estimation of the third-order aperture measures per mock takes around $14$ hours on a single CPU for the non-tomographic case, while $4.5$ hours are required on $13$ cores when including all $125$ tomographic bin combinations for the 3PCF.

We produce mock covariance matrices for both a tomographic and a non-tomographic setup and display the resulting correlation matrices in \figref{fig:TakahashiCorrcoef}. In particular, we see that the third-order aperture mass is highly correlated for radial bins of a similar angular scale. Similarly, there is a significant correlation between aperture measures being computed from different tomographic bin combinations.
We will use these sample correlation matrices to assess the detection significance of the third-order aperture measures in the KiDS-1000 data.  

Besides the correlation structure, we have validated that there is no significant presence of $B$-modes in the mock catalogs and that the measurement of the $E$-mode matches the theoretical model introduced in \cite{Heydenreichetal2022}. We refer to \appref{app:TakahashiMeas} for the corresponding figure. 
\subsection{Measurement results}
In this subsection, we present our measurement of the third-order aperture measures in the KiDS-1000 data. Throughout this subsection, we make repeated use of the `chi-square' 
\begin{align}
\chi^2 = \left[\mathrm{\mathbf{m{(\Theta)}}-\mathbf{d}}\right]^T \mathrm{Cov}^{-1}\left[\mathrm{\mathbf{m{(\Theta)}}-\mathbf{d}}\right] \ ,
\end{align}
in which $\mathbf{d}$ denotes the data vector, $\mathrm{Cov}$ stands for the covariance matrix of $\mathbf{d}$ and $\mathbf{m(\Theta)}$ is a model vector described by a set of parameters $\Theta$. As the covariance matrix is itself estimated from the T17 ensemble, we need to include the Hartlap factor \citep{Anderson2003,HartlapSchneider2007} to obtain an unbiased estimate for the precision matrix $\mathrm{Cov}^{-1}$. We quote the detection significance of $\mathbf{d}$ by virtue of its $p$-value associated with the null hypothesis of a vanishing signal, $\mathbf{m(\Theta)}\equiv0$. Unless otherwise stated, we estimate the $p$-value of some statistic measured in the KiDS-1000 data by first computing the $\chi^2$  for each of the 1944 mocks and then defining $p$ as the fraction of mocks that have a $\chi^2$ smaller than the $\chi^2$ in the real data. Finally, we introduce the signal-to-noise ($\StoN$) of a data vector by following the definition of \cite{Seccoetal2022}:
\begin{align}\label{eq:SNDef}
    \StoN \equiv 
    \begin{cases}
      \sqrt{\chi^2 - N_{\mathrm{d.o.f}}} & \mathrm{if } \ \chi^2>N_{\mathrm{d.o.f}}+1\\
      \mathrm{Null} & \mathrm{else} 
      \end{cases} \ ,
\end{align}
where $N_{\mathrm{d.o.f}}$ denotes the degrees-of-freedom, which for our setup is equal to the dimension of the data vector $\mathrm{\mathbf{d}}$. For a thorough cosmological analysis of the measurements, we refer to \cite{Burgeretal2023}.
\subsubsection{Non-tomographic measurement}
\begin{figure*}
  \centering
  \includegraphics[width=.49\textwidth,valign=t]{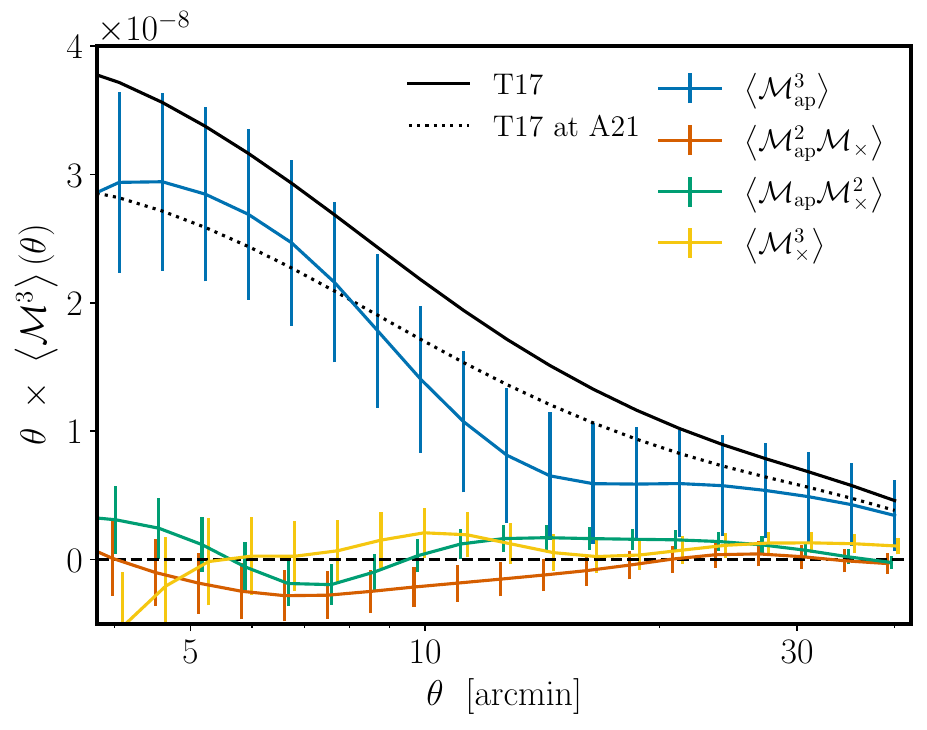}
  \includegraphics[width=.49\textwidth,valign=t]{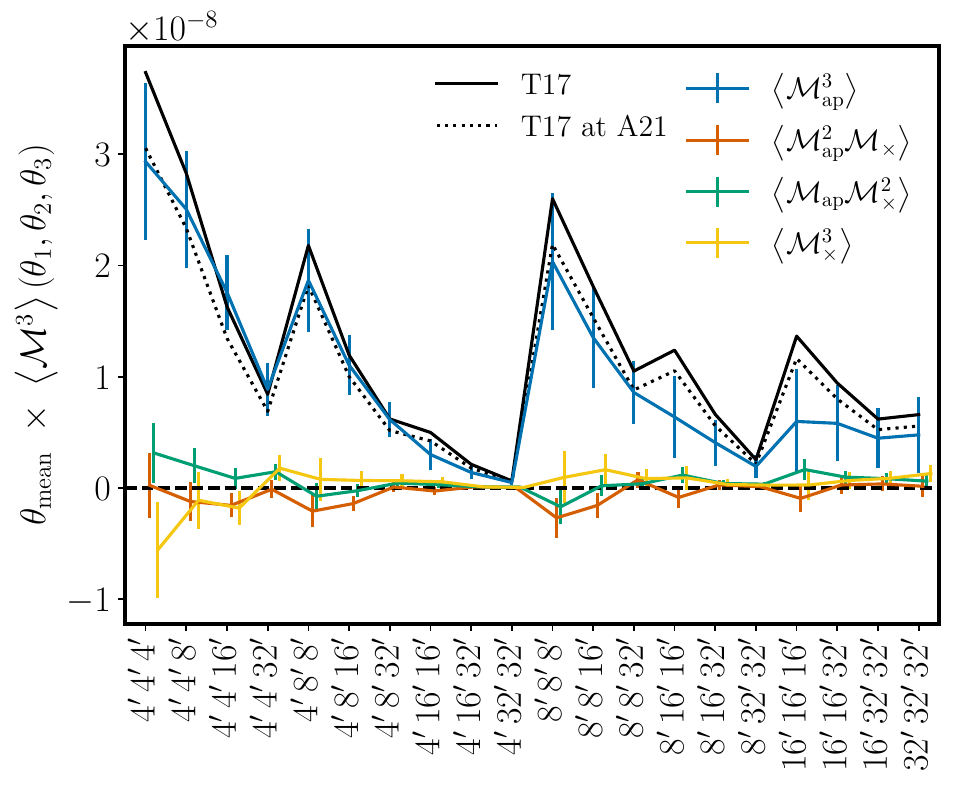}
\caption{Non-tomographic third-order aperture measures in the KiDS-1000 data, using only equal aperture scales (left) or a combination of different scales (right). The error corresponds to the standard deviation in the T17 ensemble. The black solid line compares our measurement to the mean value of the T17 ensemble and the black dotted line corresponds to the result when rescaling the measurement in the T17 ensemble to a cosmology for which the values of $\Omegam$ and $S_8$ correspond to the best-fit values from \citetalias{Asgarietal2021}}
\label{fig:RealDataNoTomo}
\end{figure*}
\begin{figure*}
  \centering
  \includegraphics[width=.99\textwidth]{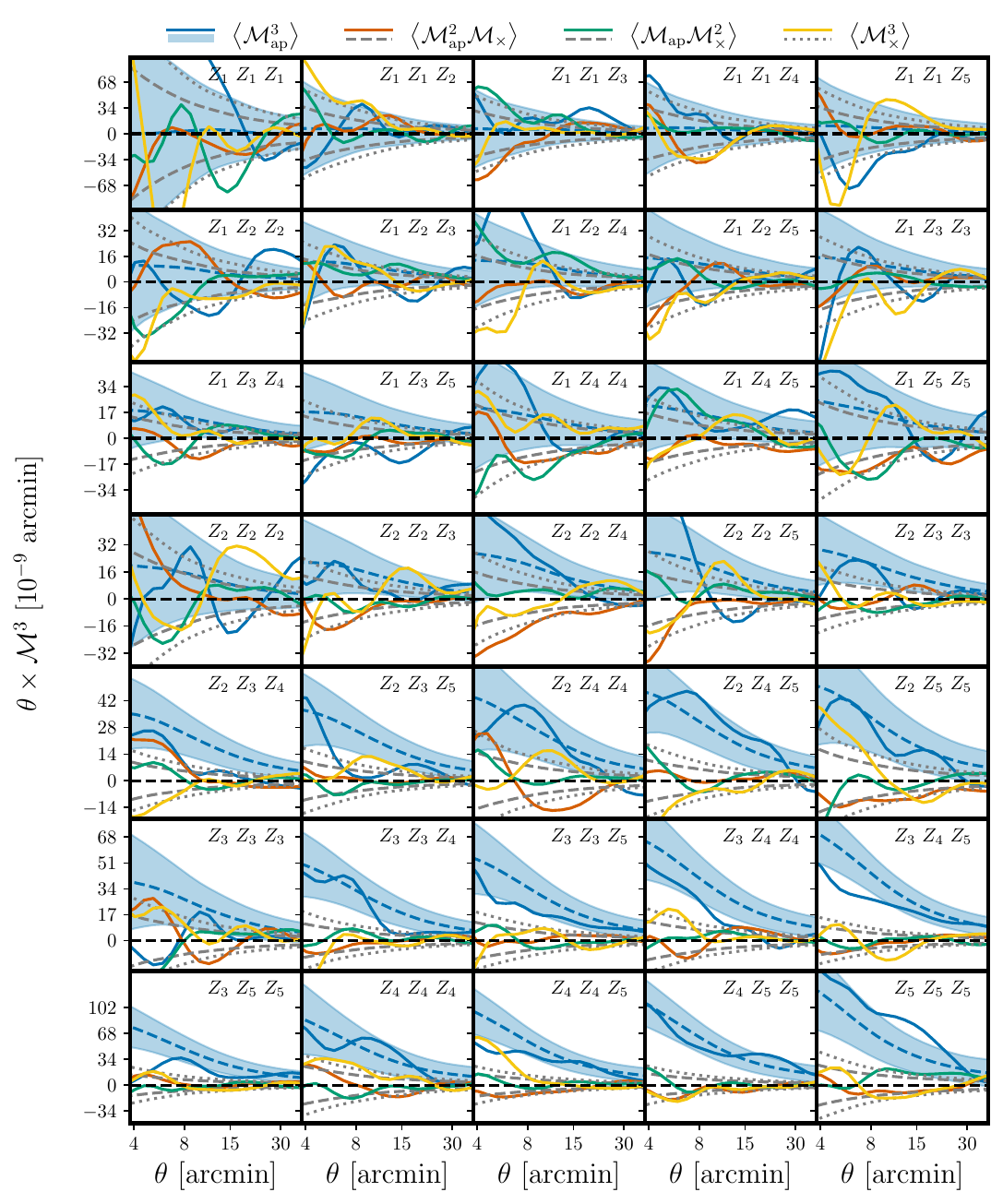}
\caption{Tomographic third-order aperture mass statistics in the KiDS-1000 data. In each panel, we show the measured statistics for some set of tomographic bins. The $\mapthree$-signal is plotted as the blue solid line while the other solid lines indicate the remaining third-order measures. We also show the mean and the $1\sigma$ error interval for the third-order aperture mass within the T17 ensemble as the blue dashed line and the blue contour. The boundaries of the  $1\sigma$ error interval for the remaining aperture measures are shown as the grey dashed and dotted curves.}
\label{fig:RealDataTomo}
\end{figure*}
\setlength{\arrayrulewidth}{0.5mm}
\setlength{\arrayrulewidth}{0.5mm}
\renewcommand{\arraystretch}{1.5}
\begin{table*}[t] 
  \centering
  \caption{Detection significance of the third-order aperture measures for a range of setups.} \label{tab:realdatadetsig}
\begin{tabular}{|c|c| *{4}{c|} *{4}{c|} *{1}{c|}}
\hline
\multirow{2}{*}{Setup} &
\multirow{2}{*}{$n_{\mathrm{data}}$} &
\multicolumn{4}{c|} {$p$-value (no cosmic variance)} &
\multicolumn{1}{c|} {$p$-value } &
\multicolumn{1}{c|} {$\StoN$ } \\
\cline{3-8}
 & & $\left\langle\mathcal{M}_{\mathrm{ap}}^3\right\rangle$ & $\left\langle\mathcal{M}_{\mathrm{ap}}^2\mathcal{M}_{\times}\right\rangle$ & $\left\langle\mathcal{M}_{\mathrm{ap}}\mathcal{M}_{\times}^2\right\rangle$ & $\left\langle\mathcal{M}_{\times}^3\right\rangle$ & $\left\langle\mathcal{M}_{\mathrm{ap}}^3\right\rangle$ & $\left\langle\mathcal{M}_{\mathrm{ap}}^3\right\rangle$ \\
\hline 
Non-tomographic (equal $\theta$) & 10 & $5.7  \times  10^{-28}$ & 0.90 & 0.09 & 0.26 & 0.05 & 3.19 \\
\hline 
Non-tomographic (mixed $\theta$) & 20 & $6.5  \times  10^{-51}$ & 0.52 & 0.27 & 0.19 & 0.02 & 4.60 \\ 
\hline 
Tomographic ($Z_1=Z_2=Z_3$) & 50 & $4.0  \times  10^{-4}$ & 0.30 & 0.08 & 0.49 & 0.17 & 3.14 \\ 
\hline 
Tomographic (all) & 350 & $9.2  \times  10^{-36}$ & 0.55 & 0.22 & 0.52 & 0.08 & 6.69 \\ 
\hline 
Tomographic ($Z_{\mathrm{min}} = 2$) & 200 & $9.8  \times  10^{-38}$ & 0.38 & 0.87 & 0.50 & 0.11 & 5.32 \\ 
\hline 
Tomographic ($Z_{\mathrm{min}} = 3$) & 100 & $1.6  \times  10^{-26}$ & 0.82 & 0.46 & 0.78 & 0.33 & 2.51 \\ 
\hline 
Tomographic ($Z_{\mathrm{min}} = 4$) & 40 & $2.9  \times  10^{-21}$ & 0.86 & 0.32 & 0.63 & 0.10 & 3.57 \\ 
\hline 
Tomographic ($Z_{\mathrm{min}} = 5$) & 10 & $1.6  \times  10^{-9}$ & 0.61 & 0.30 & 0.42 & 0.01 & 3.83 \\ 
\hline
\end{tabular} 
\vspace{.1cm} 
\tablefoot{The second column displays the length of the data vector for each setup. The next four columns display the $p$-values for the aperture measures under the assumption of no cosmological signal; in this case, the covariance for $\left\langle\mathcal{M}_{\mathrm{ap}}^3\right\rangle$ does not include the cosmic variance contribution. In the final two columns, we show the $p$-value and the $\StoN$ of the $\left\langle\mathcal{M}_{\mathrm{ap}}^3\right\rangle$ data vector when including cosmic variance. For all but the second row only equal-scale apertures are used.}
\end{table*} 
We estimate the non-tomographic third-order aperture measures in the KiDS-1000 data with the same angular binning that was used to obtain the corresponding covariance matrix in the T17 ensemble. While the figures are done using the fine-binning set, all statistical tests are performed on the coarsely binned set of aperture radii. We have further repeated the measurement using various different setups like a finer angular binning with the combined estimator, using only the discrete estimator, excluding the patch-overlap and a run using \treecorr. In all cases, we find that the measurements scatter around each other at a level of $\sim\! 10 \% $ of the surveys' error budget at scales $\lesssim 20'$, while for larger scales, the scatter increases a bit between the estimates utilizing the overlap between footprints and the estimators that do not. All conclusions drawn from each of the setups are the same as the ones presented below.

In \figref{fig:RealDataNoTomo}, we show our measurement of the non-tomographic third-order aperture measures, out of which only $\mapthreeens$ seems to be significantly different from zero. For testing the null hypothesis `no cosmological $\mapthreeens$ signal', we use the $p$-value introduced above but calculate the $\chi^2$ using a covariance matrix that is consistent with a noise-only signal (i.e. the $\mxthreeens$ covariance matrix). In this case, we compute the $p$-value using the percentile-point function of the $\chi^2$ distribution, which yields a value of $p=5.7 \times 10^{-28}$, strongly rejecting the null hypothesis. Repeating this hypothesis test (but computing the $p$-value as described before Eq. \ref{eq:SNDef})  
for the remaining three modes, we find the two parity modes and the cross mode have $p$-values that are consistent with a non-detection, i.e. $p=0.90$ for $\maptwomxens$, $p=0.26$ for $\mxthreeens$, and $p=0.09$ for $\mapmxtwoens$. Besides the detection significance in the noise-only case, we can also compute the $p$-value and the signal-to-noise of $\mapthreeens$, for which we find $p=0.05$ and $\StoN=3.19$. We list these values, as well as the values for all other setups described in this and the following subsection in Table \ref{tab:realdatadetsig}.

When comparing the measured $\mapthreeens$ to the mean value obtained from the T17 ensemble, we find that, while the measured $\mapthree$ follows a similar shape, it also has a substantially lower amplitude compared to the T17 ensemble\footnote{We note that in the KiDS-North patch there is a relatively strong depletion of the signal on intermediate aperture scales $\theta \in [10',20,]$. This also translates to the analysis of the full data set, resulting in the dip-like feature on those aperture scales visible in \figref{fig:RealDataNoTomo}.}. The majority of this discrepancy is explained by the value of $\sigma_8=0.82$ in the T17 ensemble, which is substantially higher than the best-fit values of $\sigma_8$ measured in the KiDS-1000 data at the two-point level \citep[][hereafter A21]{Asgarietal2021}. Further, as the T17 ensemble was constructed from a set of discrete lens planes in the T17, the $n(z)$ between the T17 ensemble and the KiDS-1000 data does slightly differ. To quantify the magnitude of these two effects we compute the theoretical prediction using the model of \cite{Heydenreichetal2022} for the T17 ensemble, as well as for a setup using the KiDS-1000 $n(z)$ and the best-fit values for $\Omega_m\equiv 0.245$ and $S_8\equiv 0.759$ from the KiDS-1000 analysis presented in \citetalias{Asgarietal2021}. We then multiply the measurement in the T17 ensemble with the ratio between both theory predictions and find the resulting theoretical prediction to be in agreement with our measurement in the KiDS-1000 data ($p$=0.54). 

For the measurement of the unequal-scale statistics displayed in the right panel of \figref{fig:RealDataNoTomo}, we normalize the signal by $\theta_{\mathrm{mean}}\equiv\left(\theta_1 \, \theta_2 \, \theta_3\right)^{1/3}$ and we see similar features as in the measured equal-scale statistics, namely a slightly lower amplitude of the signal compared to the T17 ensemble. Again, we can strongly reject the null hypothesis of no cosmological $\mapthreeens$ signal, and we also do not report a significant detection of the parity-violating modes and the cross-mode. Additionally, we see that the generalized aperture statistics has a significantly higher signal-to-noise than the equal-scale statistics $(\StoN=4.60)$. However, we note that this does not imply much tighter cosmological constraints as the $\StoN$ measure does not take into account the dependence of the measurement on cosmological parameters. Indeed, \cite{Heydenreichetal2022} have demonstrated that only little cosmological information is gained by including unequal aperture scales in a non-tomographic analysis and a similar test for a tomographic setup is carried out in \cite{Burgeretal2023}, arriving at the same conclusions. Again, our measurement can be well described by the model evaluated at the \citetalias{Asgarietal2021} cosmology ($p=0.48$).
\subsubsection{Tomographic measurement}
Similar to the non-tomographic case, we present the measurement carried out using the same setup as for the covariance matrix. Again, choosing a different setup for the 3PCF computation, we draw the same conclusions. Due to computational constraints, we perform the measurement using $\treecorr$ only on a few tomographic bin combinations, all of which yield consistent results. 

Our measurement results for the $35$ non-redundant tomographic bin combinations are shown in \figref{fig:RealDataTomo}. Besides the measurement, we also show the statistical uncertainty for each aperture measure, where we note that the uncertainties for $\maptwomxens$ and $\mapmxtwoens$ are equal to each other. As expected, the $\mapthreeens$ measurements are noise-dominated for combinations of low-redshift bins, while for higher mean redshifts, the signal is significantly different from zero.

For assessing the detection significance of the aperture measures in the tomographic setup we look at the whole data vector but also perform several splits to assess potential sources of redshift-dependent systematics. In particular, we choose a data vector that only considers the auto-tomographic bin combinations, as well as a set of data vectors for which the lowest redshift bin is $Z_{\mathrm{min}} \  \in \{2,3,4,5\}$. For all cases, we significantly rule out the null hypothesis of a non-detection of $\mapthreeens$, and we also do not find any significant signal for any of the remaining three measures. We refer to Table \ref{tab:realdatadetsig} for corresponding values.
\subsubsection{Comparison to previous work}
It is instructive to qualitatively compare our measurement to some previous results in the literature. \cite{Fuetal2014} have measured the non-tomographic third-order aperture statistics in the CFHTLenS data \citep{Heymansetal2012} consisting of $4.2$ million galaxies and a cosmological analysis has been carried out in \cite{Simonetal2015}. Compared to our results, they do find their measured $\mapthreeens$ signal to be in rough agreement with their mock catalogs on scales $\lesssim \! 10'$. Although the error budget of the CFHTLenS survey is substantially larger than the one of KiDS-1000, they also observe a dip-like feature of the $\mapthreeens$ signal on larger scales. They further report a significant level of $\mapmxtwoens$ in their measurement. 

The most recent measurement of $\mapthreeens$ has been carried out on the Y3 ellipticity catalog of the Dark Energy Survey, consisting of roughly $100$ million galaxies \citep{Seccoetal2022}. When comparing their non-tomographic measurement to a noiseless mock catalog constructed from one of the T17 simulations, they do not observe a dip-like feature but find a strongly depleted signal of $\mapthree$ relative to the mock on aperture scales of less than $10'$. While they do not detect a significant value of $\mapmxtwoens$, one can see in their Fig. 3 a similar behaviour of the $\mapmxtwoens$ mode on scales between $10'$ and $20'$ as in our measurement. 

\section{Conclusions}\label{sec:Conclusions}
In this work, we presented and validated a novel estimator for the natural components of the third-order shear correlation functions and applied it to the KiDS-1000 data. 

Our estimator extends the work of \cite{SlepianEisenstein2016}, and it builds on a multipole decomposition of the shear 3PCF for which a quadratic time complexity can be achieved when projecting the natural components of the shear 3PCF in a specific way, which we dub the $\times$-basis. We further introduced a grid-based approximation, in which the ellipticity catalog is mapped onto a regular grid, for which the 3PCF multipoles can be computed via FFT methods. We then merged both prescriptions in the `combined estimator' that uses the exact computation for small angular scales, while for larger scales the grid-based approximation is employed. After including the impact of non-uniformly distributed data in the estimator, we extended the previously derived equations to a tomographic setup. We then validated our estimator and checked the applicability of the grid-based approximation on the SLICS simulation suite, for which we found that the grid-based approximation can be safely used for angular separations of at least $20$ times the pixel scale. We furthermore compared our estimator against the state-of-the-art estimation method \treecorr and found an excellent agreement between both approaches with the multipole-based approach providing a speedup of several orders of magnitude. Following a short review of how to numerically obtain the third-order aperture measures from the shear 3PCF, we validated our numerical implementation of this transformation against the direct estimator, finding a sub-percent-level agreement when using a logarithmic radial bin width $b=1.05$ for the 3PCF estimation. 

Next, we proceeded to the measurement of the third-order aperture measures in the KiDS-1000 data. We constructed a suite of $1944$ mock ellipticity catalogs from the T17 ray-tracing simulations from which we computed a sample covariance matrix that includes the effect of the KiDS-1000 footprint, the angular distribution of the galaxies, as well as the correlation between the galaxy weights and the measured shapes. For the non-tomographic analysis, we reported a significant detection of a cosmological $\mapthreeens$ signal and also found that the parity-violating modes, as well as the cross mode were consistent with a non-detection. For the tomographic analysis, we again report a significant detection of the $\mapthreeens$ signal while none of the remaining three modes had a suspiciously low $p$-value. Finally, we qualitatively compared our measurement results against previous measurements in the CFHTLenS and the DES-Y3 data.

We have demonstrated that a multipole-based estimator of the shear 3PCF provides accurate results in a fraction of the time of state-of-the-art estimation procedures. This will make it feasible to utilize these methods for measuring a third-order signal in forthcoming stage-IV surveys and for constructing an accurate numerical covariance matrix taking into account all survey specifics. Extensions of these methods to higher orders are conceptually straightforward though technically cumbersome, and we leave them to future work. 

%
%

\begin{acknowledgements}
This paper went through the KiDS review process, where we want to thank the KiDS internal reviewer and Joachim Harnois-D{\'e}raps for useful comments.
  The figures in this work were created with {\sc matplotlib} \citep{Hunter2007}
  Some of the results in this paper have been derived using the healpy and HEALPix\footnote{currently http://healpix.sourceforge.net} package \citep{Gorskietal2005, Zonca2019}. We further make use of the {\sc numpy} \citep{Harrisetal2020} and {\sc scipy} \citep{Virtanenetal2020} software packages. LP acknowledges support from the DLR grant 50QE2002. SH is supported by the U.D Department of Energy, Office of Science, Office of High Energy Physics under Award Number DE-SC0019301. LL is supported by the Austrian Science Fund (FWF) [ESP 357-N]. We acknowledge use of the lux supercomputer at UC Santa Cruz, funded by NSF MRI grant AST 1828315.
\\
Author contributions: All authors contributed to the development and writing of this paper. 
\\
Some results in this paper are based on observations made with ESO Telescopes at the La Silla Paranal Observatory under programme IDs 177.A-3016, 177.A-3017, 177.A-3018 and 179.A-2004, and on data products produced by the KiDS consortium. The KiDS production team acknowledges support from: Deutsche Forschungsgemeinschaft, ERC, NOVA and NWO-M grants; Target; the University of Padova, and the University Federico II (Naples).
  
\end{acknowledgements}

%
%

\bibliography{bibliography.bib}

%

\begin{appendix}
\appendix
\section{Additional implementation specifics and scaling}\label{app:EstimatorComplexity}
\begin{figure}
  \centering
  \includegraphics[width=.99\linewidth]{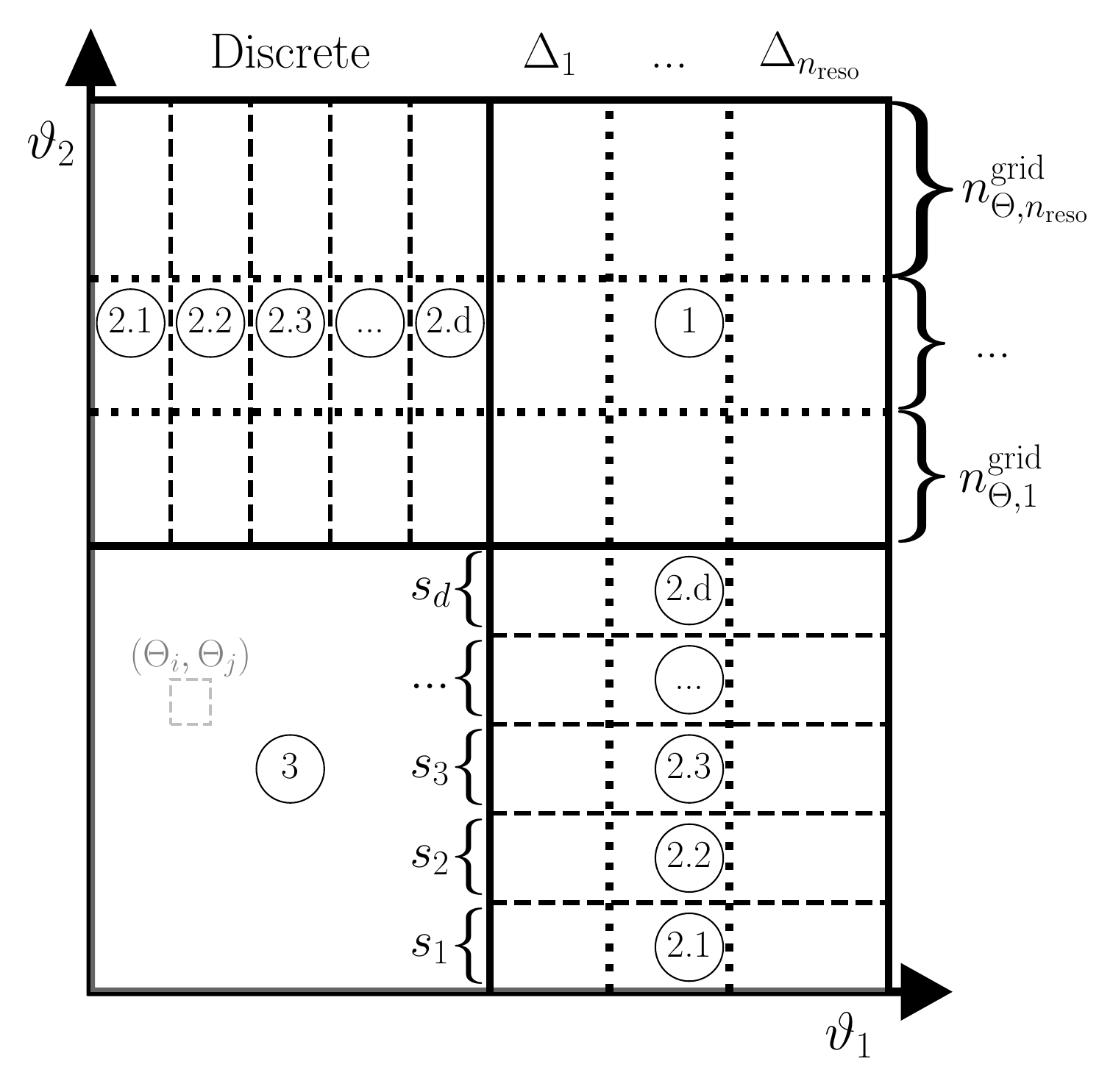}
\caption{Decomposition of the $(\vartheta_1,\vartheta_2)$-plane used for our implementation of the combined estimator. The grid itself is tiled into $n_{\Theta}^2$ elements, each corresponding to a combination of angular bins $(\Theta_i,\Theta_j)$ and each of those elements falls into one of the blocks \protect\circled{\small{1}}, \protect\circled{\small{2.i}} or \protect\circled{\small{3}}.}
\label{fig:CombinedDecomposition}
\end{figure}
\subsection{Notation and high-level overview}
In this appendix we look at the time- and space complexity of the combined estimator. For the remainder of this section, we assume a patch of $N_\mathrm{g}$ galaxies sampled at density $\overline{n}$ across $n_\mathrm{z}$ tomographic bins and we furthermore assume that we are interested in measuring all multipole components such that we can evaluate the 3PCF \Eqbr{eq:GammaMultipolesWithEC} between the zeroth and the $n_\mathrm{max}$th multipole on $n_\Theta$ radial bins. We additionally assume that we choose a strategy in which we compute the first (smallest) $n_\Theta^{\mathrm{disc}}$ radial bins using the discrete estimator while the remaining  $n_\Theta^{\mathrm{grid}}\equiv n_\Theta-n_\Theta^{\mathrm{disc}}$ radial bins are computed using the grid-based estimator with $n_{\mathrm{reso}}$ resolutions $\Delta_i\equiv2^{i-1}\Delta_1$ for which the original ellipticity catalog is mapped to $N_{\mathrm{pix}}^{(\Delta_i)}$ nonempty pixels and for each resolution scale $n_{\Theta,i}^{\mathrm{grid}}$ radial bins are used, such that $\sum_{i=1}^{n_\mathrm{reso}} n_{\Theta,i}^{\mathrm{grid}} = n_\Theta^{\mathrm{grid}}$. For later convenience, we also define a partition of the set of discrete radii into sets of $({s}_1, \cdots {s}_d)$ subintervals of size ${s}_i$.

To ensure a good compromise between the time- and space complexity of the estimator, we decompose the $(\Theta_1, \Theta_2)$-plane into several blocks as shown in \figref{fig:CombinedDecomposition} and calculate them in the following order: For the first block (gg-block) \circled{\small{1}} we calculate the multipoles in which both $\Theta$-bins are computed using the grid-based approximation. We then compute the second blocks (gd-blocks) \circled{\small{2}} that consist of the multipoles for which only one of the $\Theta$-bins is computed using the grid-based approximation while the other one uses the discrete method. Since in most applications, this step consumes the most memory we subdivide this block into $d$ sub-blocks \circled{\small{2.i}} only using ${s}_i$ elements from the discrete bins. Finally, for the third block (dd-block) \circled{\small{3.1}} we compute the remaining multipoles using the discrete prescription and then allocate the last gd-block \circled{\small{2.d}}. 
\subsection{Time complexity}
For step \circled{\small{1}} the time complexity is of order\footnote{In this subsection we only mention contributions to the time- and space complexity that can become dominant for a given survey and measurement setup.}
\begin{align}\label{eq:TimeComplexity1}
    &n_{\mathrm{max}} \ \  n_z \ \  \sum_{i=1}^{n_{\mathrm{reso}}} n_{\Theta,i}^{\mathrm{grid}} \ \mathcal{O}\left[N_{\mathrm{pix}}^{(\Delta_i)} \ \log\left(N_{\mathrm{pix}}^{(\Delta_i)}\right)\right] \nonumber \\
    &+ \ \ n_{\mathrm{max}} \ \ n_z^3 \ \   \sum_{i=1}^{n_{\mathrm{reso}}} \left\{\left[\left(n_{\Theta,i}^{\mathrm{grid}}\right)^2 + 2\left(n_{\Theta,i}^{\mathrm{grid}} \sum_{j=1}^{i-1} n_{\Theta,j}^{\mathrm{grid}}\right)\right] \ \mathcal{O}\left(N_{\mathrm{pix}}^{(\Delta_i)}\right)\right\} \ ,
\end{align}
where the terms in the first line correspond to the computation of the $G_n^{(\Delta_i)}$ via FFTs and the second line concerns the time taken to average them to obtain the multipole components. In particular, the first term in the second line considers the blocks in which the resolution scales are equal while the second term in the second line runs over all remaining terms contained in \circled{\small{1}}. Noting that $N_{\mathrm{pix}}^{(\Delta_{i+1})} \approx 
0.25 \, N_{\mathrm{pix}}^{(\Delta_i)}$, these expressions show the time benefit of using coarser grid resolutions for large radial separations.\footnote{In our implementation the prefactor is slightly larger than $0.25$ as for each resolution scale we filter out empty pixels, guaranteeing that $N_{\mathrm{pix}}^{(\Delta_1)} \leq N_{\mathrm{gal}}$ for any resolution scale.} We note that both terms scale differently in the number of radial and tomographic bins; while for a coarse radial binning and a non-tomographic setup, the first term dominates, this is not true in general when the additional prefactors in the second term surpass the additional logarithmic contribution in the first term.

The time complexity for each of the \circled{\small{2.i}} is
\begin{align}\label{eq:scalingstep2i}
    n_{\mathrm{max}} \ \overline{n} &\left(\Theta_{{s}_i,\mathrm{up}}^2 - \Theta_{{s}_i,\mathrm{low}}^2\right) \ \mathcal{O}\left(N_{\mathrm{g}}\right) 
  + \nonumber \\
    &n_{\mathrm{max}} \ \  n_z^3 \ \  {s}_i  \ \  \sum_{i=1}^{n_{\mathrm{reso}}} n_{\Theta,i}^{\mathrm{grid}} 
 \ \mathcal{O}\left( N_{\mathrm{pix}}^{(\Delta_i)}\right) \ ,
\end{align}
where both lines correspond to the same operations as the ones in \Eq{eq:TimeComplexity1}. We note that the $n_{\mathrm{max}}$ factor in the first line is well approximated with unity as in the construction of the $G_n^{\mathrm{disc}}$ in \Eq{eq:Gn_disc} only requires a single evaluation of the complex exponential from which all values for $n$ can be inferred by a few addition and multiplication operations. We further note that between two different sub-intervals ${s}_i$, ${s}_j$ both terms consist of fully complementary computations and that the second line effectively scales as if the ellipticities were distributed on regular grids.

For the first part of the final step \circled{3} we have
\begin{align}\label{eq:scalingstep3a}
    n_{\mathrm{max}} \ \left[ \ \left(\overline{n} \ \Theta_{{s}_d,\mathrm{up}}^2\right) 
    \ \ \ + \ \ \ 
    n_z^2  \ \  \left(n_{\Theta}^{\mathrm{disc}}\right)^2 \right] \ \mathcal{O}\left(N_{\mathrm{g}}\right)
\end{align}
in which both terms can be interpreted similarly to the scaling of \circled{2.i}, but the second term is scaling only quadratically in terms of the number of $z$-bins, as one would expect from \Eq{eq:tomoGn}.  

The computation of the final gd-block \circled{2.d} is equal to the formal scaling of the second line in \Eq{eq:scalingstep2i}. We note that the first term in \Eq{eq:scalingstep3a} is the only step in the method for which calculations are repeated, namely the allocation of the $G_n^{\mathrm{disc}}$ in the interval $[\Theta_{{s}_0,\mathrm{low}},\Theta_{{s}_d,\mathrm{low}}]$. For most realistic cases $\Theta_{{s}_d,\mathrm{low}} \lesssim 0.5  \, \Theta_{{s}_d,\mathrm{up}}$, meaning that the repeated calculations only constitute a subdominant fraction of the time complexity.

\subsection{Space complexity}
From the observation that the time complexity of the computation of the $G_n^{\mathrm{disc}}$ is dominated by the computation of the first value $G_1$, we aim to allocate the $\Upsilon^\times_{\mu,n}$ for all multipole orders at once. This implies that for the allocation of the gd-blocks one needs to keep all relevant  $G_n^{\Delta_i}$ in memory - the corresponding space complexity is of order
\begin{align}
    n_{\mathrm{max}} \ \ n_z \ \ \sum_{i=1}^{n_{\mathrm{reso}}} n_{\Theta,i}^{\mathrm{grid}} \ \mathcal{O}\left( N_{\mathrm{pix}}^{(\Delta_i)}\right) \ .
\end{align}
Additionally, from \Eq{eq:CombinedEstimatorRegridding} follows the requirement to map the $w\gammac \ G_n^{\mathrm{disc}}$ to grids of the corresponding pixel resolutions which results in a memory footprint of order
\begin{align}
    n_{\mathrm{max}} \ n_z^2 \ \ n_{\Theta}^{\mathrm{disc}} \ \ \sum_{i=1}^{n_{\mathrm{reso}}} \mathcal{O}\left( N_{\mathrm{pix}}^{(\Delta_i)}\right) \ .
\end{align}
In almost all cases, this term has the largest memory contribution\footnote{For example, for $n_{\mathrm{max}}=30$, $n_z=5$, $n_{\Theta}^{\mathrm{disc}}=20$ and a cumulative sum of reduced pixels $\sum_{i=1}^d N_{\mathrm{pix}}^{(\Delta_i)}$ of order $10^6$ one finds, after restoring constant prefactors, a memory requirement of $\approx 4.7 \times 10^9$ complex-valued entries.}, which lead to our subdivision of this step into $d$ substeps only requiring $1/d$-th part of the memory while leaving the time complexity nearly untouched. 

Finally, the storage requirement of the computed 3PCF multipoles is $5\, (n_\mathrm{max}+1) \ n_\Theta^2 \ n_z^3$ complex-valued entries, where we made use of the symmetries
\begin{align}
    \Upsilon^\times_{0,-n}(\Theta_i,\Theta_j; Z_k, Z_l, Z_m) &=
    \Upsilon^\times_{0,n}(\Theta_j,\Theta_i; Z_k, Z_m, Z_l) \ , \nonumber \\
    \Upsilon^\times_{1,-n}(\Theta_i,\Theta_j; Z_k, Z_l, Z_m) &=
    \Upsilon^\times_{1,n}(\Theta_j,\Theta_i; Z_k, Z_m, Z_l) \ , \nonumber \\
    \Upsilon^\times_{2,-n}(\Theta_i,\Theta_j; Z_k, Z_l, Z_m) &=
    \Upsilon^\times_{3,n}(\Theta_j,\Theta_i; Z_k, Z_m, Z_l) \ ,  \\
    \Upsilon^\times_{3,-n}(\Theta_i,\Theta_j; Z_k, Z_l, Z_m) &=
    \Upsilon^\times_{2,n}(\Theta_j,\Theta_i; Z_k, Z_m, Z_l) \ , \nonumber \\
    \mathcal{N}_{-n}(\Theta_i,\Theta_j; Z_k, Z_l, Z_m) &=
\mathcal{N}_n\left(\Theta_j,\Theta_i; Z_k, Z_m, Z_l\right)\ \nonumber ,
\end{align}
implying that only multipoles of order $n \geq 0$ need to be allocated. For our covariance setup in the T17 ensemble with $n_{\mathrm{max}}=10$, $n_\Theta=38$ and $n_z=5$ this yields a file size of about $150\,$MB.
\section{Estimation on the celestial sphere}
\label{app:EstimatorOnSphere}

\begin{figure}
  \centering
  \includegraphics[width=.99\linewidth]{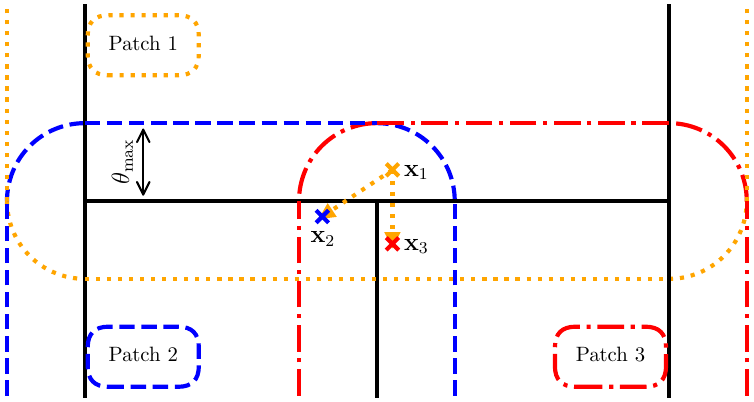}
\caption{Counting of galaxy triplets in overlapping (extended) patches. We consider a mock footprint that is split in three patches, which are extended by some scale $\theta_{\mathrm{max}}$ over their boundary. We also show three galaxies that each lie in a different patch but are contained in the intersection of the three extended patches. When computing the 3PCF multipoles in the first patch, the form of \Eq{eq:PatchGamma} only allows us to compute the $G_n$s at position $\mathbf{x}_1$, meaning that only the depicted triangle is included in the 3PCF. When encountering the same three galaxies in the two other patches, a different reference will be chosen, such that each particular triangle configuration will be six-fold counted across the full footprint.}
\label{fig:Patchoverlap}
\end{figure}
While the discussion in the main text did solely focus on the efficient estimation of the shear 3PCF on a flat geometry, an actual survey will observe galaxies on the sky and will construct its ellipticity catalog $\mathcal{F}$ based on a spherical topology. However, by cutting out small patches of the curved-sky catalog we can use the flat-sky approximation and apply the methods introduced in \secref{sec:Estimator} to compute the shear 3PCF for each patch and then average over the individual measurements to obtain an estimate for the shear 3PCF of the full data set.

In a more algorithmic fashion, for mapping a catalog on the sphere to its flat-sky analog $\mathcal{A}$, one can proceed according to the following steps:
\begin{enumerate}
    \item Define a domain decomposition of the curved-sky catalog $\mathcal{F}$ into $N_\mathrm{p}$ smaller patches\footnote{In order to minimize the curved-sky effects we construct the individual patches to have a rectangular shape with each side having a length of $\leq 10^\circ$.}
    \item Rotate each patch to the equator. In order to account for the different reference directions of the shapes we make use of the \texttt{angle\_ref} method of the \texttt{healpy.Rotator} class.
    \item Map each rotated patch onto the flat sky using a sinusoidal projection; call these patches $\mathcal{A}_i \ : \ i\in I \equiv \{1,\cdots,N_\mathrm{p}\}$
    \item Define the union of the rotated and projected patches in step 3 as the flat-sky analog of the original catalog: \\$\mathcal{F} \sim \mathcal{A} \equiv \bigcup_{i \in I}\mathcal{A}_i$
\end{enumerate}
We are now in a position to use the methods developed in \secref{sec:Estimator} 
to measure the correlators on the full survey footprint ${}^{\mathcal{A}}\Upsilon^{\times}_{\mu}$ up until some largest separation $\theta_{\mathrm{max}}$. For this we combine the multipole components ${}^{\mathcal{A}_i}\Upsilon^{\times}_{\mu,n}$ in the individual patches $\mathcal{A}_i$ as follows: 
\begin{align}\label{eq:PatchGamma}
    {}^{\mathcal{A}}\Upsilon_{0,n}^{\times}(\Theta_1,\Theta_2)
    \nonumber \\ &\hspace{-1cm}\equiv \sum_{i=1}^{\Ngal} w_i  \gammacs{,i} \ G_{n-3}^{\mathrm{disc}}(\btheta_i;\Theta_1) \ G_{-n-3}^{\mathrm{disc}}(\btheta_i;\Theta_2) 
    \nonumber \\ &\hspace{-1cm} = \sum_{p=1}^{N_{\mathrm{p}}} \sum_{i=1}^{N_{\mathrm{gal},p}} w_i  \gammacs{,i} \ G_{n-3}^{\mathrm{disc}}(\btheta_i;\Theta_1) \ G_{-n-3}^{\mathrm{disc}}(\btheta_i;\Theta_2) 
    \nonumber \\ &\hspace{-1cm} \equiv \sum_{p=1}^{N_{\mathrm{p}}} {}^{\mathcal{A}_p}\Upsilon_{\mu,n}^{\times}(\Theta_1,\Theta_2) \ .
\end{align}
We note that in the second step, the inner sum only runs over the $N_{\mathrm{gal},p}$ galaxies contained in the $\mathcal{A}_p$, but the $G_n^{\mathrm{disc}}$ are able to access data from $\mathcal{A}$ - this implicit overlap of patches is necessary to account for triplets split between different patches.
For implementing the overlap in practice, all that needs to be done is to extend the patches to also include galaxies being separated by at most $\theta_{\mathrm{max}}$ from the interior of each original patch, but to only consider galaxies within the original patch when evaluating the inner sum in \Eq{eq:PatchGamma};  see \figref{fig:Patchoverlap} for an illustration. 

\section{Transformation between shear 3PCF and third-order aperture measures}\label{app:3PCFTrafo}
In this appendix we collect the main equations that are necessary for converting a the shear 3PCF to third-order aperture measures.
\subsection{Filter functions}
An explicit form for the filter functions $F_0$ and $F_1$ has been derived in \citetalias{Schneideretal2005}. Here we add the equations for $F_2$ and $F_3$ and adapt all equations to our triangle convention. We begin by defining the helper quantities 
\begin{align}\label{eq:3pcftomap3_helpers}
    \Theta^2 &= \sqrt{\frac{\theta_1^2 \theta_2^2 + \theta_1^2 \theta_3^2 + \theta_2^2 \theta_3^2}{3}} 
    \ ,\\
    S &= \frac{\theta_1^2 \, \theta_2^2 \, \theta_3^2}{\Theta^6}
    \ , \\
    M &= S \ \frac{\vartheta_1 \ \vartheta_2}{2 \pi \, \Theta^4}
    \ , \\
    Z &=\frac{\sum_{i=1}^3(-\theta_i^2 + 2\theta_{i+1}^2 + 2\theta_{i+2}^2)|\mathbf{\breve{\mathitq_i}}|^2}{6\Theta^4} 
    \ , \\
    f_i &= \frac{\theta_{i+1}^2+\theta_{i+2}^2}{2\Theta^2} + \frac{\cqi^*\left( \cqii - \cqiii \right)}{|\cqi|^2} \ \frac{\theta_{i+1}^2-\theta_{i+2}^2}{6\Theta^2}
    \ , \\
    g_i &= \frac{\theta_{i+1}^2 \, \theta_{i+2}^2}{\Theta^4} + \frac{\cqi^*\left( \cqii - \cqiii \right)}{|\cqi|^2} \ \frac{\theta_i^2\left(\theta_{i+2}^2-\theta_{i+1}^2\right)}{3\Theta^4}
    \ ,
\end{align}
where $i \in \{1,2,3\}$ and all indices should be taken $\mathrm{mod} \, 3$. We then have for the filter functions $F_\mu$:
\begin{align}
    F_0(\vartheta_1,&\vartheta_2,\phi; \theta_1,\theta_2,\theta_3) 
    \nonumber \\ &= 
    \frac{M}{24} \ \frac{|\cqa|^2|\cqb|^2|\cqc|^2}{\Theta^6} \, f_1^{*\, 2} f_2^{*\, 2} f_3^{*\, 2} \ \ee^{-Z} \ , \\
    F_1(\vartheta_1,&\vartheta_2,\phi; \theta_1,\theta_2,\theta_3) 
    \nonumber \\ &= 
    M \, \ee^{-Z} \left[ \frac{|\cqa|^2|\cqb|^2|\cqc|^2}{24\Theta^6}f_3^{*\, 2} f_1^{*\, 2} f_2^{2} \right. 
    \nonumber \\  &\hspace{1.2cm}-\frac{1}{9}\frac{\cqc\cqa\cqb^{*\,2}}{\Theta^4} f_3^{*} f_1^{*} f_2 \, g_2^* 
    \nonumber \\  &\hspace{1.2cm}
    +\frac{1}{27}\left(\frac{\cqc^2 \cqa^2 \cqb^{*\,4}}{|\cqa|^2|\cqb|^2|\cqc|^2 \, \Theta^2} \, g_2^{* \, 2}\right. 
    \nonumber \\  &\hspace{2.0cm}
    \left.\left. + \frac{2\theta_3^2\theta_1^2}{\Theta^4} \ \frac{\cqc\cqa\cqb^{*\,2}}{|\cqb|^2\Theta^2} \, f_3^* f_1^* \right)\right] \ ,
\end{align}
where the $\cqi$ are given in \eqref{eq:QvecDef}. The other two components follow by cyclically permuting the indices in $F_1$, i.e. $F_2 = F_1^{(231)}$ and $F_3 = F_1^{(312)}$. When comparing to the expressions in \citetalias{Schneideretal2005}, their $F_i$ is our $F_{i+2}$.
\subsection{Transformation to the aperture measures}
By evaluating the expressions $\expval{\mathcal{M}^3}$, $\expval{\mathcal{M}^*\mathcal{M}^2}$, $\expval{\mathcal{M} \mathcal{M}^*\mathcal{M}}$ and $\expval{\mathcal{M}^2 \mathcal{M}^*}$ of the complex aperture measures in terms of the (cross) aperture mass via $\expval{\mathcal{M}} = \expval{\mathcal{M}_{\mathrm{ap}}} + \ii \expval{\mathcal{M}_{\times}}$ one can invert the corresponding linear system to obtain the transformation equations
{\allowdisplaybreaks
\begin{align}
    &\mapthreeens(\theta_1,\theta_2,\theta_3) \nonumber \\
&\quad \quad =\Re\Large[\expval{\mathcal{M}^2 \mathcal{M}^*}(\theta_1,\theta_2,\theta_3)+\expval{\mathcal{M} \mathcal{M}^* \mathcal{M}}(\theta_1,\theta_2,\theta_3) \nonumber \label{eq:MapTrafo8_1} \\
& \quad \quad \quad +\expval{\mathcal{M}^*\mathcal{M}^2}(\theta_1,\theta_2,\theta_3)+\expval{\mathcal{M}^3}(\theta_1,\theta_2,\theta_3)\Large]/4 
\ , \\
&\left\langle\mperp \map^2 \right\rangle(\theta_1,\theta_2,\theta_3) \nonumber \nonumber \\
&\quad  = \Im\Large[\expval{\mathcal{M}^2 \mathcal{M}^*}(\theta_1,\theta_2,\theta_3)+\expval{\mathcal{M} \mathcal{M}^* \mathcal{M}}(\theta_1,\theta_2,\theta_3) \nonumber \\
& \quad  \quad - \expval{\mathcal{M}^*\mathcal{M}^2}(\theta_1,\theta_2,\theta_3)+\expval{\mathcal{M}^3}(\theta_1,\theta_2,\theta_3)\Large]/4 
\ , \\
&\left\langle\map \mperp \map \right\rangle(\theta_1,\theta_2,\theta_3) \nonumber \\
&\quad  = \Im\Large[\expval{\mathcal{M}^2 \mathcal{M}^*}(\theta_1,\theta_2,\theta_3)-\expval{\mathcal{M} \mathcal{M}^* \mathcal{M}}(\theta_1,\theta_2,\theta_3) \nonumber \\
& \quad  \quad +\expval{\mathcal{M}^*\mathcal{M}^2}(\theta_1,\theta_2,\theta_3)+\expval{\mathcal{M}^3}(\theta_1,\theta_2,\theta_3)\Large]/4 
\ , \\
&\left\langle\map^2 \mperp \right\rangle(\theta_1,\theta_2,\theta_3) \nonumber \\
&\quad  = \Im\Large[-\expval{\mathcal{M}^2 \mathcal{M}^*}(\theta_1,\theta_2,\theta_3)+\expval{\mathcal{M} \mathcal{M}^* \mathcal{M}}(\theta_1,\theta_2,\theta_3) \nonumber \\
& \quad  \quad + \expval{\mathcal{M}^*\mathcal{M}^2}(\theta_1,\theta_2,\theta_3)+\expval{\mathcal{M}^3}(\theta_1,\theta_2,\theta_3)\Large]/4 
\ , \\
&\left\langle\map \mperp^2 \right\rangle(\theta_1,\theta_2,\theta_3) \nonumber \\
&\quad  = \Re\Large[\expval{\mathcal{M}^2 \mathcal{M}^*}(\theta_1,\theta_2,\theta_3)+\expval{\mathcal{M} \mathcal{M}^* \mathcal{M}}(\theta_1,\theta_2,\theta_3) \nonumber \\
& \quad  \quad -\expval{\mathcal{M}^*\mathcal{M}^2}(\theta_1,\theta_2,\theta_3)-\expval{\mathcal{M}^3}(\theta_1,\theta_2,\theta_3)\Large]/4 
\ , \\
&\left\langle\mperp \map \mperp \right\rangle(\theta_1,\theta_2,\theta_3) \nonumber \\
&\quad  = \Re\Large[\expval{\mathcal{M}^2 \mathcal{M}^*}(\theta_1,\theta_2,\theta_3)-\expval{\mathcal{M} \mathcal{M}^* \mathcal{M}}(\theta_1,\theta_2,\theta_3) \nonumber \\
& \quad  \quad +\expval{\mathcal{M}^*\mathcal{M}^2}(\theta_1,\theta_2,\theta_3)-\expval{\mathcal{M}^3}(\theta_1,\theta_2,\theta_3)\Large]/4 
\ , \\
&\left\langle\mperp^2 \map \right\rangle(\theta_1,\theta_2,\theta_3) \nonumber \\
&\quad  = \Re\Large[-\expval{\mathcal{M}^2 \mathcal{M}^*}(\theta_1,\theta_2,\theta_3)+\expval{\mathcal{M} \mathcal{M}^* \mathcal{M}}(\theta_1,\theta_2,\theta_3) \nonumber \\
& \quad  \quad +\expval{\mathcal{M}^*\mathcal{M}^2}(\theta_1,\theta_2,\theta_3)-\expval{\mathcal{M}^3}(\theta_1,\theta_2,\theta_3)\Large]/4 
\ , \\
&\left\langle\mperp^3 \right\rangle(\theta_1,\theta_2,\theta_3) \nonumber \\
&\quad  = \Im\Large[\expval{\mathcal{M}^2 \mathcal{M}^*}(\theta_1,\theta_2,\theta_3)+\expval{\mathcal{M} \mathcal{M}^* \mathcal{M}}(\theta_1,\theta_2,\theta_3) \nonumber \\
& \quad  \quad +\expval{\mathcal{M}^*\mathcal{M}^2}(\theta_1,\theta_2,\theta_3)-\expval{\mathcal{M}^3}(\theta_1,\theta_2,\theta_3)\Large]/4 
\ .\label{eq:MapTrafo8_8}
\end{align}
}
In the main text we average over the three permutations of the correlators where both, $\map$ and $\mperp$ are present. This implies that the covariance of those components is smaller than the covariance of $\left\langle\mperp^3 \right\rangle$.

For a tomographic setup we need to make the adjustment that each correlator also depends on the redshift bins, i.e. we have
\begin{align}
    \mapthreeens(\theta_1,\theta_2,\theta_3) \rightarrow \mapthreeens(\theta_1,\theta_2,\theta_3;Z_1,Z_2,Z_3) \ .
\end{align}
We note that due to the symmetries of the form
\begin{align}
    \mapthreeens(&\theta_1,\theta_2,\theta_3;Z_1,Z_2,Z_3) 
    \nonumber \\ &= \mapthreeens(\theta_1,\theta_2,\theta_3;Z_{\pi(1)},Z_{\pi(2)},Z_{\pi(3)})
    \ , \\
    \left\langle\mperp^2\map\right\rangle (&\theta_1,\theta_2,\theta_3;Z_1,Z_2,Z_3) \nonumber \\ &= \left\langle\map\mperp^2\right\rangle (\theta_3,\theta_1,\theta_2;Z_1,Z_2,Z_3) \ , 
\end{align}
for any $\pi \in S_3$, the transformation equations can be reduced to the shorter expressions found i.e. in \citetalias{Schneideretal2005}:
{\allowdisplaybreaks
\begin{align}
&\mapthreeens(\theta_1,\theta_2,\theta_3) \nonumber \\
&\quad  =\Re\Large[
\expval{\mathcal{M}^*\mathcal{M}^2}(\theta_1,\theta_2,\theta_3)+
\expval{\mathcal{M}^*\mathcal{M}^2}(\theta_1,\theta_3,\theta_2) \nonumber \\
& \quad \quad  +\expval{\mathcal{M}^*\mathcal{M}^2}(\theta_2,\theta_3,\theta_1)+\expval{\mathcal{M}^3}(\theta_1,\theta_2,\theta_3)\Large]/4 
\ , \label{eq:MapTrafo4_1} \\ 
&\left\langle\map^2 \mperp \right\rangle(\theta_1,\theta_2,\theta_3) \nonumber \\
&\quad  =\Im\Large[-
\expval{\mathcal{M}^*\mathcal{M}^2}(\theta_1,\theta_2,\theta_3)+
\expval{\mathcal{M}^*\mathcal{M}^2}(\theta_1,\theta_3,\theta_2) \nonumber \\
& \quad \quad  +\expval{\mathcal{M}^*\mathcal{M}^2}(\theta_2,\theta_3,\theta_1)+\expval{\mathcal{M}^3}(\theta_1,\theta_2,\theta_3)\Large]/4 
\ , \\ 
&\left\langle\map \mperp^2 \right\rangle(\theta_1,\theta_2,\theta_3) \nonumber \\
&\quad  =\Re\Large[
\expval{\mathcal{M}^*\mathcal{M}^2}(\theta_1,\theta_2,\theta_3)+
\expval{\mathcal{M}^*\mathcal{M}^2}(\theta_1,\theta_3,\theta_2) \nonumber \\
& \quad \quad  -\expval{\mathcal{M}^*\mathcal{M}^2}(\theta_2,\theta_3,\theta_1)-\expval{\mathcal{M}^3}(\theta_1,\theta_2,\theta_3)\Large]/4 
\ , \\ 
&\left\langle\mperp^3 \right\rangle(\theta_1,\theta_2,\theta_3) \nonumber \\
&\quad  =\Im\Large[
\expval{\mathcal{M}^*\mathcal{M}^2}(\theta_1,\theta_2,\theta_3)+
\expval{\mathcal{M}^*\mathcal{M}^2}(\theta_1,\theta_3,\theta_2) \nonumber \\
& \quad \quad  +\expval{\mathcal{M}^*\mathcal{M}^2}(\theta_2,\theta_3,\theta_1)-\expval{\mathcal{M}^3}(\theta_1,\theta_2,\theta_3)\Large]/4 
\ , \label{eq:MapTrafo4_4}
\end{align}
}
where we suppressed the redshift bin
indices for better readability. 

We note that if one is solely interested in the values of $\mapthreeens$ and $\left\langle\mperp^3\right\rangle$, as well as in the averaged quantities
\begin{align}
    \left\langle\map\mperp^2\right\rangle^{\mathrm{av.}} &\equiv 
    \frac{\left( 
    \left\langle \mperp^2\map\right\rangle +
    \left\langle \mperp \map \mperp \right\rangle +
    \left\langle \map \mperp^2 \right\rangle
    \right)}{3} \ , \\
     \left\langle\map^2\mperp\right\rangle^{\mathrm{av.}} &\equiv 
    \frac{\left( 
    \left\langle \map^2\mperp\right\rangle +
    \left\langle \map \mperp \map \right\rangle +
    \left\langle \mperp \map^2 \right\rangle
    \right)}{3} \ ,
\end{align}
it is sufficient to only compute the 3PCF for $Z_1 \leq Z_2 \leq Z_3$, if one uses the full transformation equations (\ref{eq:MapTrafo8_1}-\ref{eq:MapTrafo8_8}) and additionally symmetrizes them over the three aperture radii. 

\section{Measurements in the T17 Ensemble}\label{app:TakahashiMeas}
\begin{figure*}
  \centering
  \includegraphics[width=.99\textwidth]{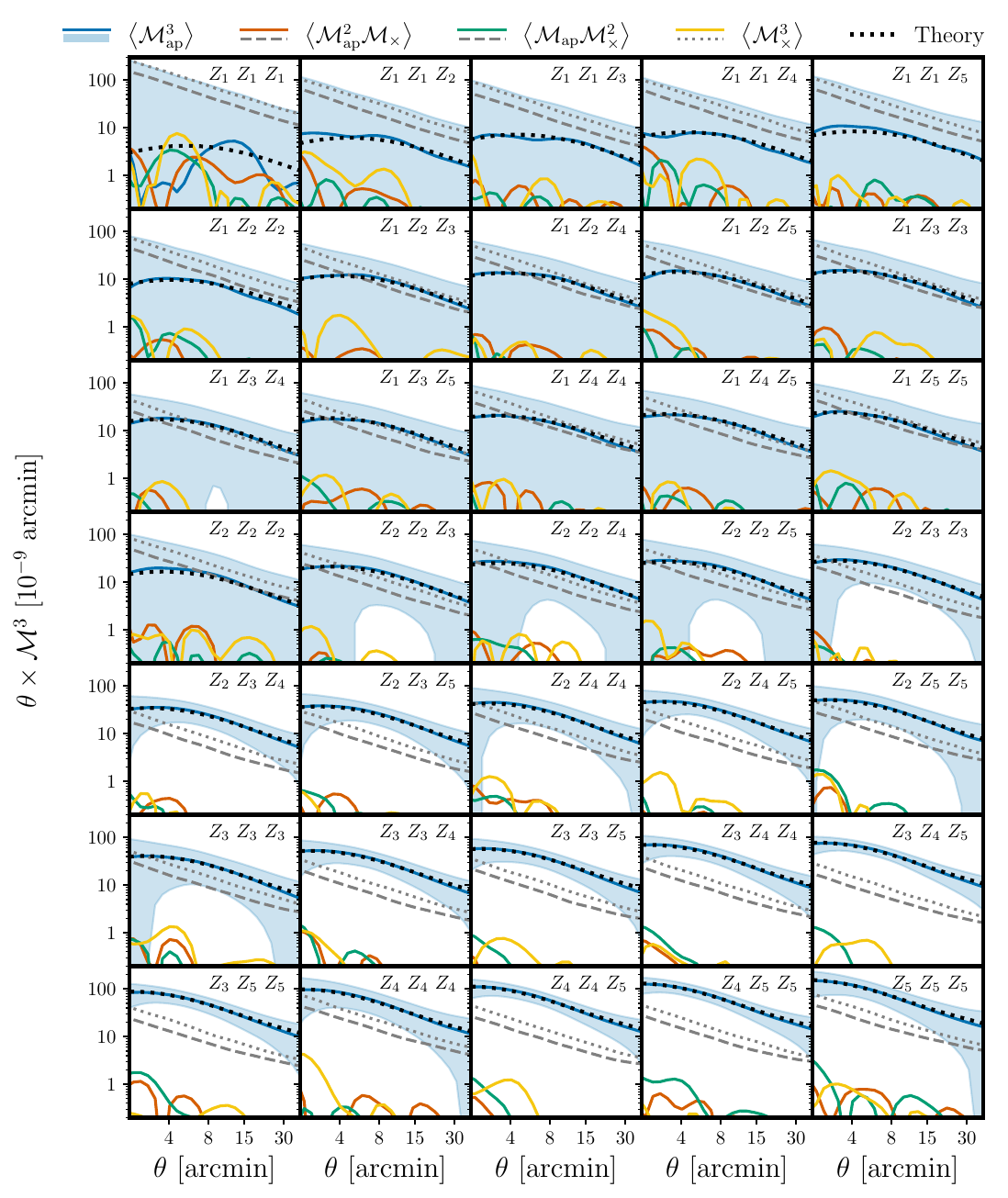}
\caption{Tomographic third-order aperture mass statistics in the T17 ensemble. In each panel, we show the measured statistic for one configuration of the tomographic bins. The mean $\mapthree$ signal and its standard deviation are plotted as the blue solid line and the blue error band. The other solid lines indicate the absolute values of the remaining third-order measures and the $1\sigma$ uncertainty level of those measures are indicated by the grey dashed (dotted) lines. The theoretical prediction is shown as the black dotted line.}\label{fig:T17DataTomo}
\end{figure*}

\figref{fig:T17DataTomo} displays the measurement of the third-order aperture statistics in the T17 ensemble consisting of $1944$ mock ellipticity catalogs mimicking the KiDS-1000 survey. We see that most of the cosmological signal is contained in tomographic bin combinations consisting of high--$z$ bins. However, for a cosmological analysis, one should also include the low--$z$ bins, as those will help constraining parameters related to the intrinsic alignment model. We also see that none of the three cross-aperture measures yields a significant signal, showing that neither such spurious signal is present in the mock ellipticity catalogs nor that it is created by our estimation procedure. 
Finally, we compare our measurement to the theoretical prediction, for which we are using the model introduced in \cite{Heydenreichetal2022}. On small angular scales, we see that the measurement does slightly over-predict the theory, which we attribute to the reduced-shear effect that are included in the T17 ensemble, but not in the theory. On large angular scales, however, the measurement does slightly underpredict the theoretical prediction. One reason for this could lie in the lens-shell thickness in the T17 ensemble; in particular \citet{Takahashietal2017} show that on the level of the shear 2PCF this effect can cause a suppression of the signal of up to $10\%$ for angular separations of about $100'$. Besides these two effects, a part of the discrepancy between the theory and the measurement could also lie in the limited accuracy of the Bihalofit model \citep{Takahashietal2020}, which is at the $10\%$-level.

\section{Accuracy requirements}
\begin{figure*}
  \centering
  \includegraphics[width=.9\textwidth]{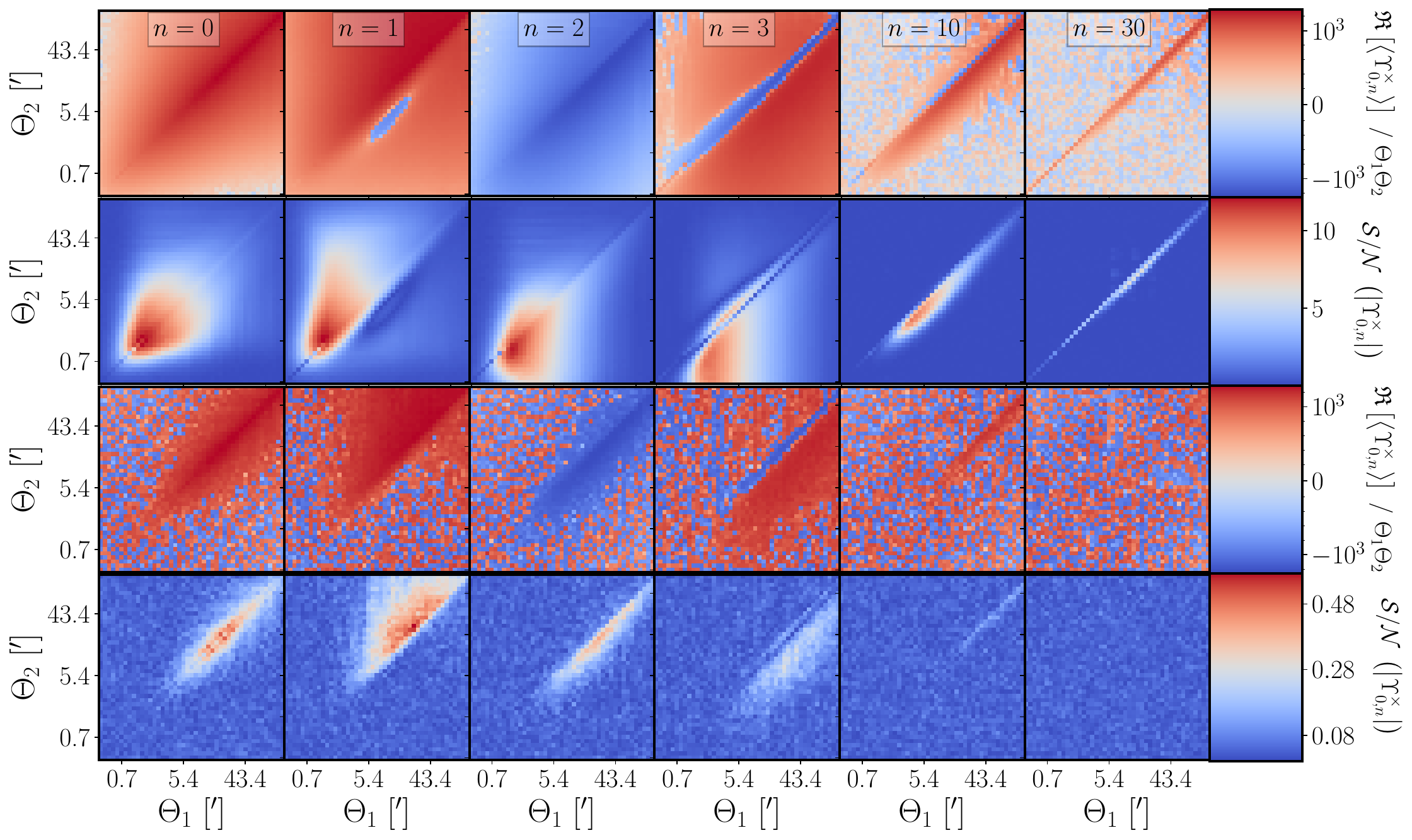}
\caption{Mean signal and detectability of the higher-order multipoles $\Upsilon_{0,n}^\times$ in the SLICS ensemble, after averaging over all 819 realizations. In the top row, we show the mean signal from the noiseless SLICS ensemble. To aid the visualization we normalize the multipoles by the product of the corresponding angular bins and further use a logarithmic scaling for positive and negative values in the colormap. In the second row, we show the signal-to-noise ratio of the modulus of the corresponding multipole component. The third and fourth rows display the same measurements when shape noise is included in the mocks.}\label{fig:Gamma0MultipolesConvergence}
\end{figure*}
\subsection{Three Point Correlation Function}\label{app:3PCFConvergence}
In \figref{fig:Gamma0MultipolesConvergence} we show the detection significance of some multipoles $\Upsilon_{0,n}^{\times}$ in the SLICS ensemble for noisy and noiseless data. We see that most of the signal contribution stems from the lowest-order multipoles. This is expected as the shear 3PCF does not contain high-frequency oscillations in the $\varphi$ direction. We also see that shape noise strongly depletes the detection significance of the individual elements; we note that for a stage-IV survey having a significantly higher source density this effect will not be as severe. The plots for the three remaining $\Upsilon_{\mu,n}^{\times}$ all follow the same trend.
\subsection{Third-order aperture mass}\label{app:Map3Convergence}
\begin{figure}
    \centering
    \includegraphics[width=.9\linewidth]{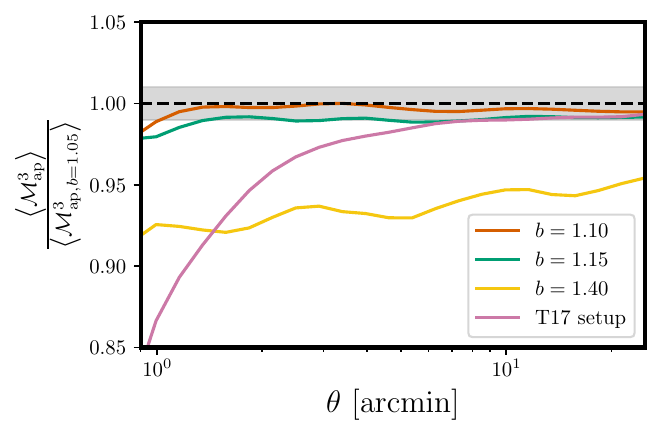}
    \includegraphics[width=.9\linewidth]{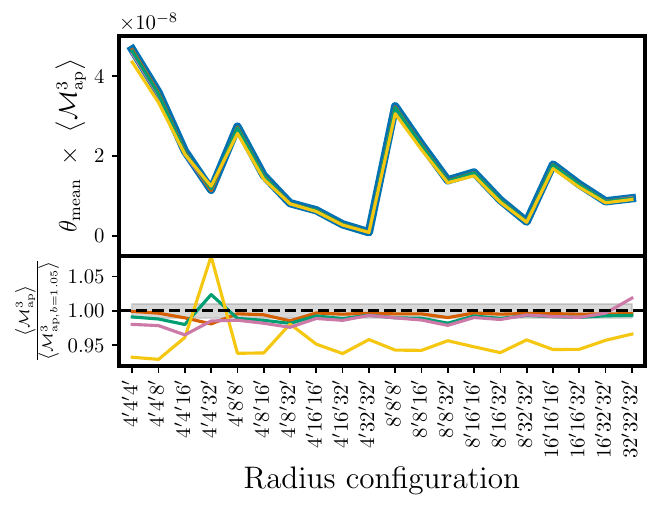}
    \caption{Accuracy of the numerical integration of the shear 3PCF to the third-order aperture statistics for different logarithmic spacings of radial bins, as well as for the radial binning that was used for the computation of the covariance in the T17 ensemble. All measurements are done in the SLICS ensemble. In the upper panel, we show the ratio between $\mapthreeens$ when estimated using some logarithmic bin width $b$ for the shear 3PCF and the corresponding result with $b \equiv 1.05$. In the lower panel, we show the corresponding results when allowing for unequal aperture filters.}
    \label{fig:Map3Convergence_nr}
\end{figure}

With the third-order aperture mass being an integrated measure of the shear 3PCF for which the time and space complexity are strongly dependent on the number of radial bins and the largest considered multipole, one should investigate the convergence of the numerical integration for various binning choices. In \figref{fig:Map3Convergence_nr} we fix the largest multipole $n_{\mathrm{max}} = 20$ and compare the numerical integration of the 3PCF for a range of logarithmic bin widths $b$. For the equal-scale apertures, we find that choosing a value of $b \approx 1.1$ yields per-cent level accuracy. This value is similar to the corresponding value found by previous studies \citep{Fuetal2014, Seccoetal2022} examining the convergence of the numerical integration using TreeCorr. We further show in \figref{fig:Map3Convergence_nr} that our choice of $b\approx 1.15$ for the computation of the numerical covariance from the T17 ensemble yields sufficiently accurate data vectors for aperture scales of more than $4'$.\footnote{We note that as the ellipticity data in the T17 ensemble itself is pixelized on a grid with a sidelength of $\approx \, 0.4'$, the measurements of $\mapthreeens$ are biased low for small aperture scales, independent of the smallest separation considered in the 3PCF.}  For the multiscale aperture statistics we find similar convergence properties, except for the cases when two aperture radii are considerably smaller than the third one.

\begin{figure}
    \centering
    \includegraphics[width=.9\linewidth]{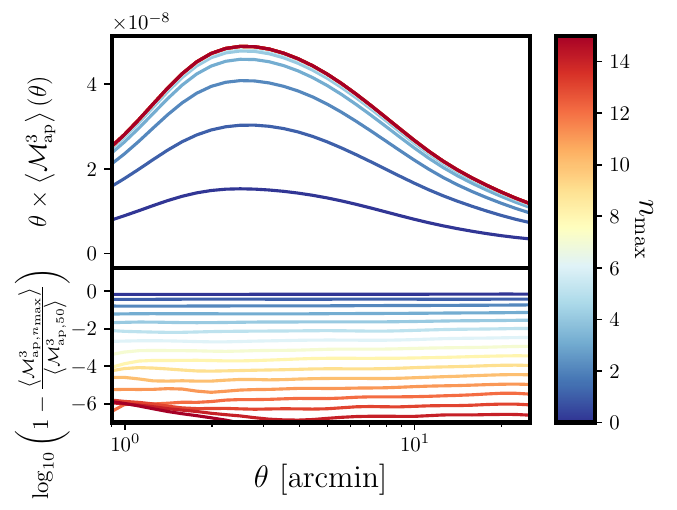}
    \includegraphics[width=.9\linewidth]{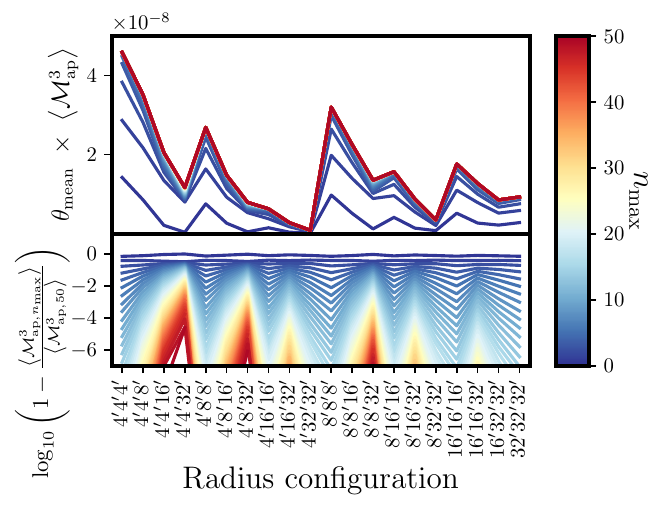}
    \caption{Impact of higher-order multipoles of the shear 3PCF on the third-order aperture statistics in the SLICS ensemble. In the upper panel of the top (bottom) figure, we show the estimated signal for the equal-scale (unequal-scale) statistics, when using only the first $n_{\mathrm{max}}$ multipoles. The lower panel displays the decimal logarithm of the fractional difference between the measurement using $50$  multipoles and the measurements using only multipoles up until $n_{\mathrm{max}}$.}
    \label{fig:Map3Convergencenmax}
\end{figure}

In \figref{fig:Map3Convergencenmax} we compare how limiting the number of considered multipoles impacts the expected value of $\mapthreeens$. We find, as before, a very fast convergence for the equal-scale aperture statistics, while for the general case, there is a strong dependence of the level of convergence on the ratio between the two smallest and the largest aperture radii. This behaviour can be understood as follows: In the case of two small and one large aperture radii the associated $\mapthreeens$ is dominated by squeezed triangle configurations. As such configurations mainly receive contributions from the multipole components $\Upsilon_{\mu,n}(\Theta, \Theta)$ and those are precisely the ones having a nonvanishing and detectable signal for large multipole orders (see \figref{fig:Gamma0MultipolesConvergence}), we do expect a slower convergence.
\end{appendix}

\end{document}